\documentclass[10pt,journal,doublecolumn]{IEEEtran}
\usepackage{graphics} 
\usepackage{epsfig} 
\usepackage{mathrsfs}
\usepackage{times} 
\usepackage{amsmath} 
\usepackage{algpseudocode}
\usepackage{amssymb}
\usepackage{algorithm}
\usepackage{flushend}
\usepackage{tabularx}
\usepackage{booktabs}
\usepackage{multirow}
\usepackage{multicol}
\usepackage{epstopdf}
\usepackage{url}
\newcommand{\figwidth}{0.9\columnwidth}
\newcommand{\figwidthh}{0.9\columnwidth}
\usepackage{tabularx,multirow}
\usepackage{wrapfig}
\usepackage{mathpazo}
\usepackage[refpage]{nomencl} 
\usepackage{rotating}
\usepackage{psfrag}
\usepackage{listings}
\usepackage{booktabs}
\usepackage{multirow,esdiff}

\usepackage{threeparttable}

\usepackage[noadjust]{cite}

\allowdisplaybreaks

\usepackage[mathscr]{eucal}

\makeatletter
\newcommand{\biggg}{\bBigg@{3}}
\newcommand{\vast}{\bBigg@{4}}
\newcommand{\Vast}{\bBigg@{5}}
\makeatother

\usepackage{tikz}
\usepackage{standalone}

\usepackage{subcaption}

\usepackage{pifont}
\newcommand{\cmark}{\ding{51}}%
\newcommand{\xmark}{\ding{55}}%

\begin{document}
%
\title{A Survey on Resource Allocation in Vehicular Networks}

\author{Md. Noor-A-Rahim,~
Zilong Liu,~
      Haeyoung Lee,~
      G. G. Md. Nawaz Ali,~
       Dirk Pesch,~
       Pei Xiao
       \thanks{Md. Noor-A-Rahim and Dirk Pesch are with the  School of Computer Science \& IT, University College Cork,  Ireland  (E-mail: {\tt \{m.rahim,d.pesch\}@cs.ucc.ie}). Zilong Liu is with School of Computer Science and Electrical Engineering, University of Essex, United Kingdom (E-mail: {\tt zilong.liu@essex.ac.uk}). Haeyoung Lee and Pei Xiao are with Institute for Communication Systems, 5G Innovation Centre, University of Surrey, United Kingdom (E-mail: {\tt \{haeyoung.lee,p.xiao\}@surrey.ac.uk}). G. G. Md. Nawaz Ali  is with the Department of Applied Computer Science, University of Charleston, USA   (E-mail: {\tt ggmdnawazali@ucwv.edu}).
}}  
\maketitle

\begin{abstract}
Vehicular networks, an enabling technology for Intelligent Transportation System (ITS), smart cities, and autonomous driving, can deliver numerous on-board data services, e.g., road-safety, easy navigation, traffic efficiency, comfort driving, infotainment, etc. Providing satisfactory Quality of Service (QoS) in vehicular networks, however, is a challenging task due to a number of limiting factors such as erroneous and congested wireless channels (due to high mobility or uncoordinated channel-access), increasingly fragmented and congested spectrum, hardware imperfections, and anticipated  growth of vehicular communication devices. Therefore, it will be critical to allocate and utilize the available wireless network resources in an ultra-efficient manner. In this paper, we present a comprehensive survey on resource allocation  schemes for the two dominant vehicular network technologies, e.g. Dedicated Short Range Communications (DSRC) and cellular based vehicular networks. We discuss the challenges and opportunities for resource allocations in modern vehicular networks and outline a number of promising future research directions.
\vspace{10pt}
\end{abstract}


\IEEEpeerreviewmaketitle

\section{Introduction}
The prevalent vision is that vehicles (e.g., cars, trucks, trains, etc.) will in the future be highly connected with the aid of ubiquitous wireless networks, anytime and anywhere, which is expected to lead to improved road safety, enhanced situational awareness, increased travel comfort, reduced traffic congestion, lower air pollution, and lower road infrastructure costs.
Central to this vision is a scalable and intelligent vehicular network which is responsible for efficient information exchange among vehicles and/or between vehicles, other road users and road side infrastructure (Vehicle-to-Everything (V2X) communications). As an instrumental enabler for Intelligent Transportation Systems (ITS), smart cities, and autonomous driving, vehicular networks have attracted significant research interests in recent years both from the academic and industrial communities \cite{Liu2017CodingAssisted,Wang2017Centrality,Nguyen2018, Cheng2015,Ali2018_tvt}.
So far, there are two major approaches for V2X communications: dedicated short range communications (DSRC) and cellular based vehicular communication \cite{Seo2016, Bazzi2017}. DSRC is supported by a family of standards including the IEEE 802.11p amendment for Wireless Access in Vehicular Environments (WAVE), the IEEE 1609.1$\sim$.4 standards for resource management, security, network service, and multi-channel operation \cite{Kenney2011DSRC}. 
On the other hand, 3GPP have been developing cellular vehicular communications, also called C-V2X, designed to operate over cellular networks such as Long-Term Evolution (LTE) and 5G new radio (5G NR). V2X allows every vehicle to communicate with different types of communication entities, such as pedestrians, Road-Side Units (RSU), satellites, internet/cloud, and other vehicles. Both V2X techniques\footnote{Besides IEEE 802.11p and 3GPP, the Internet Engineering Task Force (IETF)  has  been  working  on  V2X related  topics from a network   and   transport layer  perspective, specially making necessary changes to make IPv6 more suitable for V2X communications \cite{J.Jeongetal.2020}.} have their respective advantages and limitations when adopted in a vehicular environments. As a result, an integration into heterogeneous vehicular networks has been suggested to exploit their unique benefits, while addressing their individual drawbacks.

Wireless networks suffer from a wide range of impairments, among them shadowing, path loss, time- and/or frequency-selective wireless channels, jamming and/or multi-user interference. To deal with these impairments, radio resources (such as time slots, frequency bands, transmit power levels, etc.) should be allocated in an optimized manner to cater for varying channel and network conditions. Dynamic Resource Allocation (RA) schemes are preferred as they give rise to significantly improved performance (compared to static RA schemes) by efficiently exploiting wireless channel and network variations in a number of dimensions \cite{Georgiadis2006, Zhang2010_cog,Wang2011_mul}. For instance, authors in \cite{Botsov2014,Ren2015,Sun2016,Sun2016b,Cheng2017} studied RA schemes for  Device-to-Device (D2D) V2X networks by taking into account fast vehicular channel variations.
However, efficient resource allocation in vehicular networks is an extensive topic due to the following major challenges:
\begin{enumerate}
    \item Highly dynamic mobility scenarios covering low-speed vehicles (e.g., less than 60 km/h) to high-speed cars/trains (e.g., 500 km/h or higher)  \cite{Zhang2011,ZijunZhao2013}. The air interface design for high mobility communication, for instance, may require more time-frequency resources in order to combat the impairments incurred by Doppler spread/shifts and multi-path channels.
    \item Wide range of data services (e.g., in-car multimedia entertainment, video gaming/conferencing, ultra-reliable and low-latency delivery of safety messages, high-precision map downloading, etc) with different QoS requirements in terms of reliability, latency, and data rates. In particular, some requirements (e.g., high data throughput against ultra-reliability) may be conflicting and hence it may be difficult to support them simultaneously.
    \item Expected explosive growth of vehicular communication devices in the midst of increasingly fragmented and congested spectrum. Moreover, devices employed in vehicular networks usually have different hardware parameters and therefore may display a wide variation in their communication capabilities under different channel and network conditions. For example, a vehicular sensor device
    aiming for long battery life (e.g., more than 10 years) is unlikely to use sophisticated signal processing algorithms for power saving purposes whereas more system resources and more signal processing capabilities may be required for ultra-reliable transmission of safety messages.
\end{enumerate}

Driven by these challenges of vehicular networks but also more broadly in other types of wireless networks, a wide range of disruptive ideas and techniques for resource allocation have been published aimed at addressing various aspects of the problem space over the past decade. Many of them are covered in survey publications works addressing resource allocation in for example cognitive radio networks \cite{Naeem2014,ElTanab2017,Ahmad2015}, ultra-dense networks \cite{Teng2019}, multi-user MIMO systems \cite{Castaneda2017}. To the best of our knowledge, survey papers \cite{Harounabadi2018,Masmoudi2019} are the only ones that specifically focus on resource allocation for vehicular networks. However, while these two surveys consider resource allocation in cellular vehicular networks, they ignore resource allocation techniques for DSRC based vehicular networks. Moreover, they also do not cover more recent work such as machine learning based solutions for resource allocation in vehicular networks. To fill this gap and to stimulate further research and innovation in this area, we provide a comprehensive survey on the state-of-the-art of RA in both, DSRC and cellular vehicular networks, as well as for heterogeneous versions of these two network types. We also provide a detailed discussion on current state of machine learning based RA  and suggest a number of promising research directions.


This article is organized as follows. We start our discourse in Section II by a high-level overview of vehicular networks based on DSRC, C-V2X and heterogeneous versions. Detailed literature surveys on these three types of vehicular networks are presented in Sections III-V, respectively. As machine learning is gaining increased attention also in this paper's topic area, we provide a dedicated survey in Section VI on applications of machine learning for RA in vehicular networks. In Section VII, we summarize three important future directions for RA research in vehicular networks lead by network slicing, machine learning, and context awareness. Finally, this article is concluded in Section VIII. 

\section{Overview of Vehicular Networks}

\begin{figure*} [h]
 \centering
  \begin{subfigure}[t]{0.5\textwidth}
 \includegraphics[width=\textwidth]{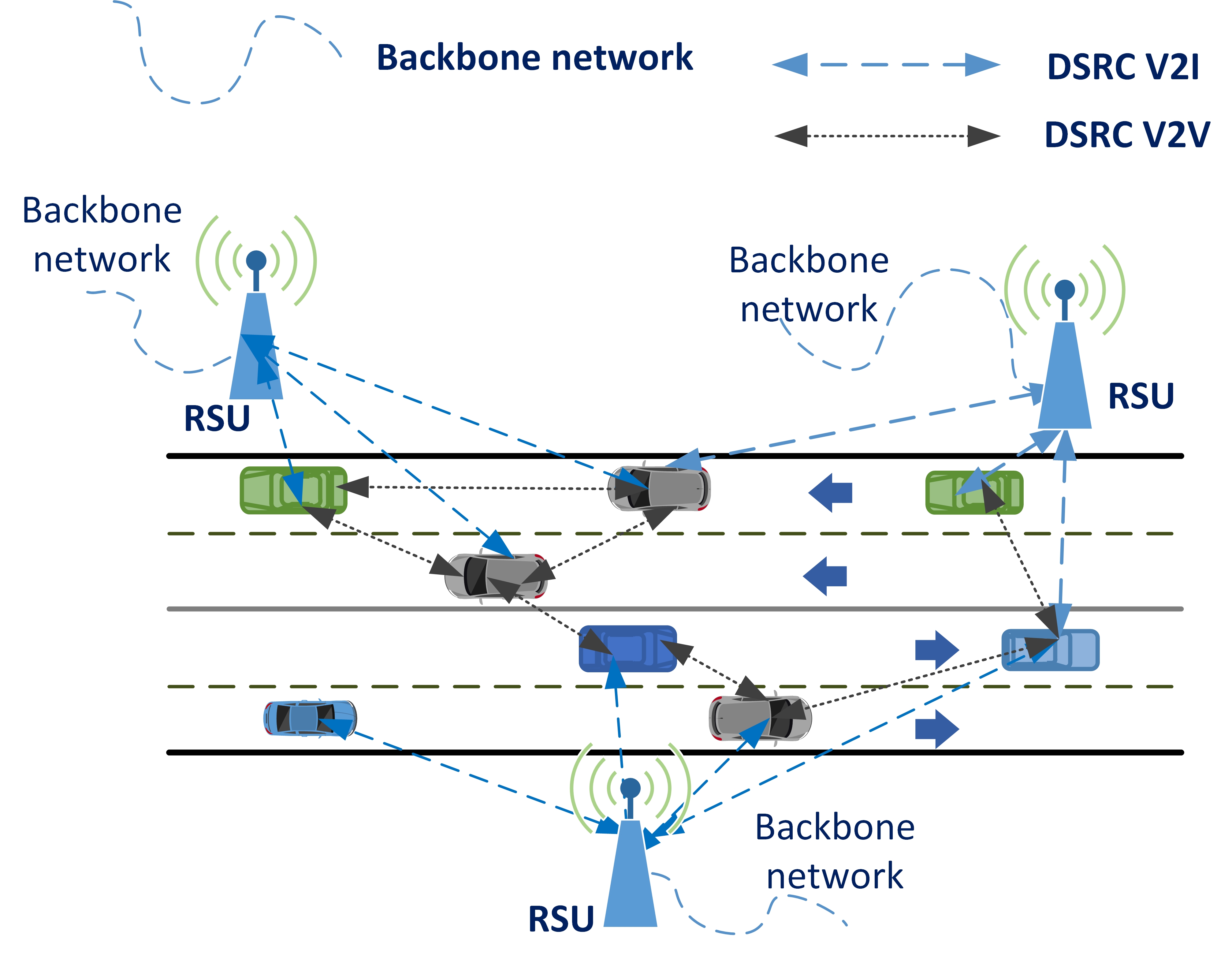}
  \caption{DSRC Vehicular Network.}
 \label{fig:DSRC}
 \end{subfigure}~
 \begin{subfigure}[t]{0.5\textwidth}
\includegraphics[width=\textwidth]{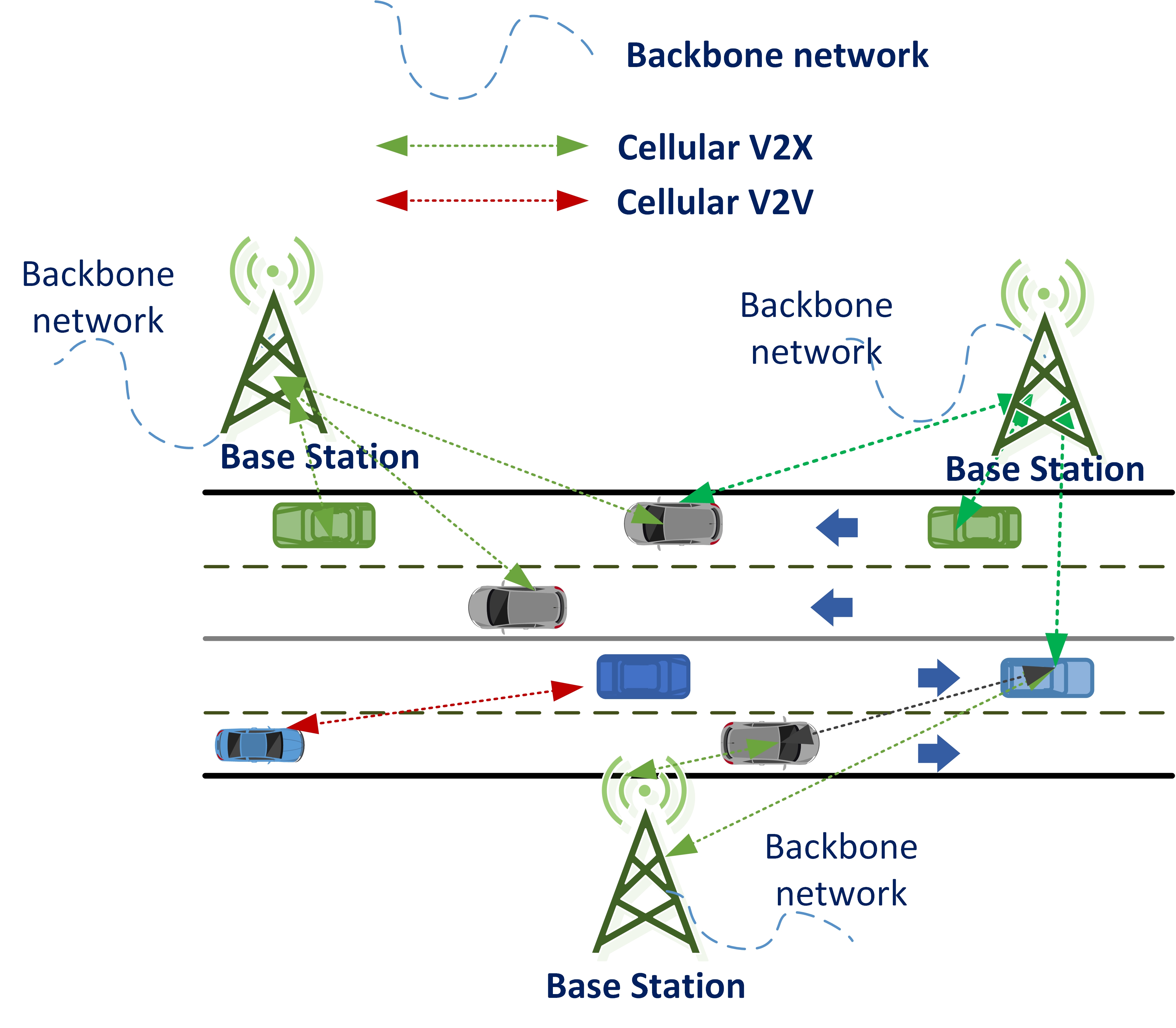}
\caption{Cellular Vehicular Network.}
\label{fig:Cellular}
 \end{subfigure}%

 \begin{subfigure}[t]{0.6\textwidth}
\includegraphics[width=\figwidthh]{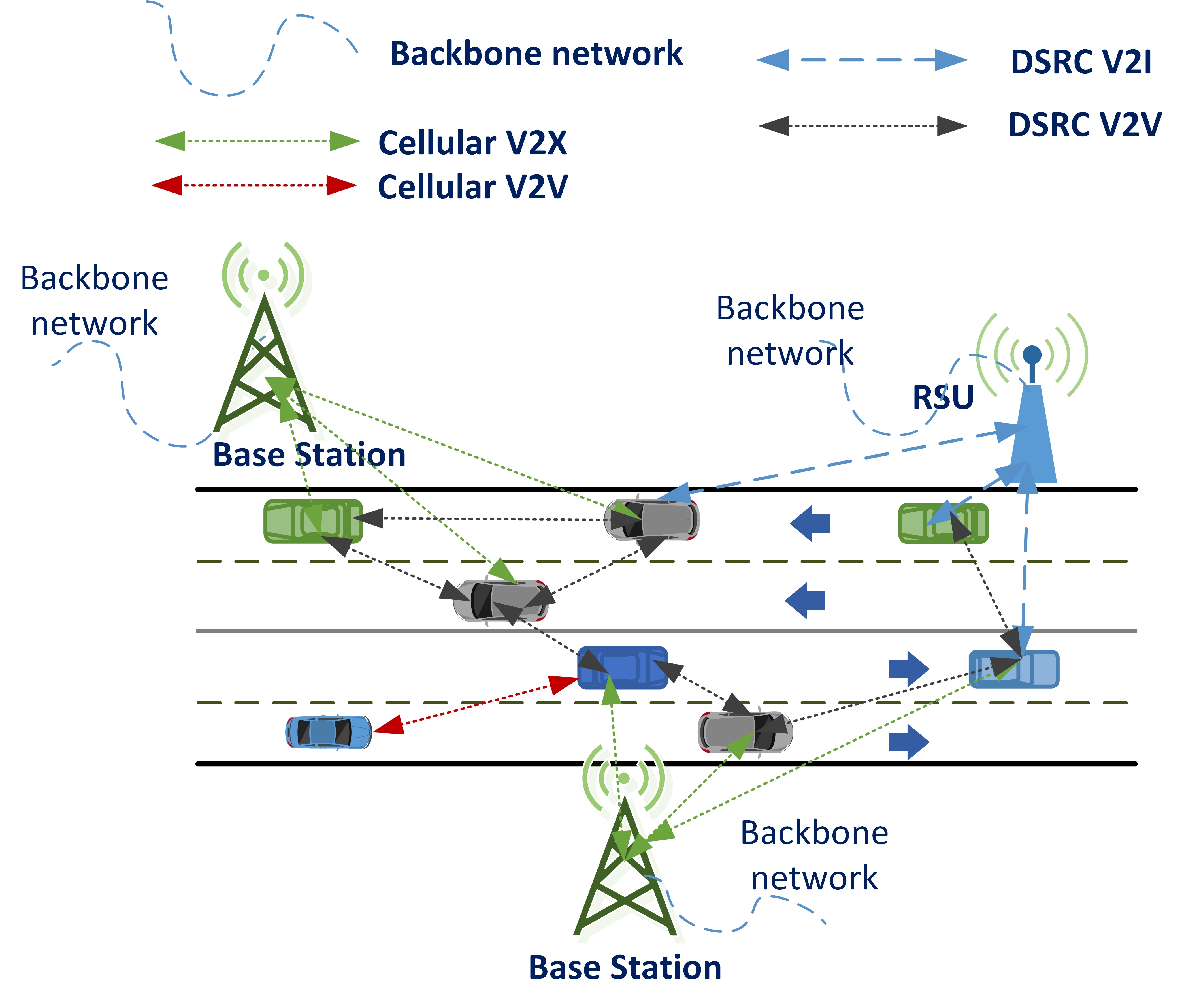}
\caption{Heterogeneous Vehicular Network.}
\label{fig:DSRCandCellular}
 \end{subfigure}%

 \caption{Overview of Vehicular Networks.}\label{fig:overview}
\end{figure*}

\subsection{DSRC Vehicular Network}
Dedicated Short Range Communications (DSRC) is a standardised  wireless technology that is designed to support ITS applications in vehicular networks. The underlying standard for DSRC is 802.11p, which is a derivative of the IEEE 802.11e with small modifications in the QoS aspects.  DSRC supports wireless communication  between vehicles and rode side units (RSUs).  The US  Department of Transportation estimates that Vehicle-to-Vehicle (V2V) communications based on DSRC can eliminate   up to 592,000 accidents involving vehicles and can save up to 1,083 lives annually in respect to crashes at intersection \cite{Nhtsa2014DSRC}. These predictions show a significant potential for the DSRC technology to reduce accidents and to improve road safety.


DSRC technology supports two  classes  of  devices \cite{Morgan2010, Hartenstein2010}: the On-Board  Unit  (OBU)  and the  Road-Side Unit (RSU), which are  equivalent  to  the  Mobile  Station  (MS)  and  Base  Station   (BS)   in   traditional cellular   systems,  respectively. An overview of a typical DSRC vehicular network in shown in Fig.~\ref{fig:DSRC}. The  Federal Communications Commission in the United States has allocated 75 MHz licensed spectrum for DSRC communications in the 5.9 GHz frequency band \cite{Noor-A-Rahim2018_acc}. Out  of  the  75  MHz  spectrum,  5  MHz  is  reserved  as  the  guard  band  and  seven  10-MHz  channels  are  defined for DSRC communications.  The  available  spectrum is configured into one Control Channel (CCH) and six Service Channels (SCHs). The  CCH  is  reserved  for  high-priority  short  messages  or  control  data,  while other data are transmitted over the SCHs. Several Modulation and Coding Schemes (MCS) are supported with the   transmitter  (TX)  power ranging  from  0  dBm  to  28.8  dBm.  Based on the communication environments, the coverage distance may range from 10m to 1km.

The fundamental mechanism  for medium/channel access in DSRC is  known as the Distributed  Coordination  Function  (DCF). With DCF, vehicles contend for a wireless channel using a Carrier-Sense Multiple Access (CSMA) with Collision Avoidance (CA) technique. To transmit a packet from a vehicle, the channel must be sensed idle for a guard period. This guard period is known as the Distributed Inter-Frame Space (DIFS). If the channel is sensed busy,  the vehicle initiates a slotted backoff process and vehicles are only permitted to start transmissions at the beginning of slots. Vehicles randomly choose their individual backoff time from the range $[0, CW-1]$, where $CW$ is known as  the contention window.  The backoff time counter is decreased by $1$, when the channel is sensed idle for a time slot. The counter is frozen when the channel is sensed occupied and reactivated after the channel is sensed idle again for a DIFS time interval period.  A vehicle transmits when its backoff counter reaches zero. A packet collision occurs when two or more vehicles choose the same time slot for transmission. Note that unlike other forms of the IEEE 802.11 standard, e.g. IEEE802.11a/b/g/n and the most recent update IEEE 802.11ax, IEEE 802.11p does not use a collision avoidance  mechanism. Consequently, DSRC networks are prone to the effects of the hidden terminal problem. Along with the above channel access mechanism, IEEE 802.11p adopts the Enhanced Distributed Channel Access (EDCA) mechanism, which allows four access categories for vehicle data transmission with different priorities.

\subsection{Cellular based Vehicular Network (C-V2X)}
Despite the fact that DSRC is generally considered the de facto standard for vehicular networks, cellular/LTE based vehicular communications (also known as C-V2X) has recently attracted significant attention due to its large coverage, high capacity, superior quality of services, and  multicast/broadcast support. An depiction of a cellular based  vehicular network is shown in Fig.~\ref{fig:Cellular}. LTE-V2V communication exploits LTE uplink resources while utilizing  Single Carrier Frequency Division Multiple Access (SC-FDMA) at  the PHY and MAC layers  \cite{Cecchini2018}.   According to the LTE specifications, the available bandwidth is subdivided into equally-spaced (spacing of 15 kHz) orthogonal subcarriers.  A Resource Block (RB) in LTE is formed by 12 consecutive subcarriers (i.e., 180 kHz) and  one time slot (i.e., 0.5 ms). The number of data bits carried by each RB depends on specific Modulation and Coding Schemes (MCS).

 To   enable   direct   short-range   communication between devices, LTE uses  direct communication interface so-called PC5 interface (also known as LTE side-link), which can be used for V2V and V2I communications.  To utilize the  available radio resources, two side-link modes are defined by the 3GPP standard release 14:  Mode 3 and Mode 4. In  Mode 3, it is assumed that the vehicles are fully covered by one or more evolved NodeBs (eNBs) who dynamically assign the resources being used for V2V communications through control signalling. This type of resource assignment is called dynamic scheduling. An eNB may also reserve a set of resources for a vehicle for its periodic transmissions. In this case, the eNB defines for how long resources will be reserved for the particular vehicle.  In  Sidelink Mode 4, vehicles are assumed to be in areas without cellular coverage and hence, resources are allocated in a  distributed manner. A sensing based semi-persistent  transmission mechanism  is introduced in Sidelink Mode 4  to enable distributed resource allocation.

The distributed algorithm
optimizes the use of the available channels by increasing the resource reuse distance  between vehicles that are using the same resources. A distributed congestion control mechanism is also applied which calculates the channel busy ratio and the channel occupancy ratio. Then, a vehicle reserves resources for a random interval and sends a reservation message, called Scheduling Assignment (SA), using Side-link Control Information (SCI). Other vehicles which sense and listen to the wireless channel find out from the SA the list of busy resources and avoid selecting those resources. To increase the reliability, a vehicle may send a data message in this mode more than once. In Release 14, 3GPP mentioned that D2D communications included in Releases 12 and 13 can also be applied to vehicular networks as the localization characteristics of vehicular networks are similar to D2D networks \cite{Lin2014,Sun2016}.

\subsection{Heterogeneous Vehicular Networks}
Despite its potential and advantages, the DSRC technology suffers from several drawbacks \cite{HameedMir2014, Seo2016,Araniti2013}, such as limited coverage, low data rate, and limited QoS guarantee, and unbounded channel access delay. These drawbacks are due to DSRC's origins in earlier IEEE 802.11 standards, which were originally designed for wireless local area networks with low mobility. Although the current DSRC technology has been shown to be effective in supporting vehicular safety applications in many field trials \cite{Araniti2013}, significant challenges remain when employing DSRC technology in some more hostile vehicular environments.

While cellular based vehicular networks can provide wide coverage and high data rate services, they may not be able to support decentralized communication as the networks may become easily overloaded in situation with very high vehicle density, e.g. traffic jams. Thus, both DSRC and cellular based vehicular networks have their respective advantages and limitations when used in vehicular environments.  A depiction of a heterogeneous vehicular network in shown in Fig.~\ref{fig:DSRCandCellular}. A range of efforts \cite{Zheng2015a, Dressler2014, Atat2012, Huang2010, LiuFuqiang2010, Zheng2015, Dai2018, Cespedes2015, Shafiee2011, He2016} have been made towards the integration of both DSRC and cellular based vehicular networks (e.g., LTE) for enhanced vehicular communications. Besides the integration of DSRC and cellular based vehicular networks, emerging V2X applications require efficient utilization of heterogeneous access technologies, such as Wi-Fi and TV broadcasting networks.

\begin{table*}[h]
\centering   

\caption{Existing RA techniques for DSRC vehicular network.}  \label{tab:comp}
\scriptsize
\begin{tabular}{| p{1cm} | p{2cm}| p{1.5cm}| p{2.5cm} |p{2cm}|p{1.5cm}|p{1cm}|p{1cm}|c|c|c|}    
\hline 
Reference  &   Scenario   & Use Case & Allocation Technique & Constraints & Optimizing parameters & Mobility  & Priority classes\\[0.5ex]  
\hline
\cite{Harigovindan2012} & Multi-lane Highway & Generic & Packet collision modelling & Fairness of channel access & Contention window & \cmark &  \cmark \\
\hline
\cite{Karamad2008} & Single-lane Highway & Generic & Throughput modelling & Residence time, Network fairness & Contention window &  \cmark &  \cmark \\
\hline
\cite{Rossi2015,Rossi2017}   & Single-lane Highway & Safety message & Connectivity and throughput modelling & Interference, velocity  &  Throughput &  \cmark  &  \xmark \\
\hline
 \cite{Alasmary2012}  & Multi-lane Highway & Safety message & Mobility based  access modelling & Mobility & Backoff mechanism &  \cmark  &  \cmark \\
\hline
\cite{Ryu2011}   & Single-lane Highway & Emergency  message & Priority based  allocation & Delay & Bandwidth & \cmark  &  \xmark  \\
\hline
\cite{Sheu2010}   & Single-lane Highway & Generic & Throughput  fairness modelling & Transmission distance   & Throughput &  \xmark  &  \xmark \\
\hline
 \cite{Ali2018}   &  Urban grid layout & Caching & Exhaustive Search & Residence time, deadline & Data rate &  \cmark  &  \xmark \\
\hline

\end{tabular}

\end{table*}

\section{Resource Allocation in DSRC Networks}
In this section, we review resource allocation approaches for DSRC based vehicular networks, which have largely focused on MAC parameter allocation, channel allocation and rate allocation techniques. In the following, we classify resource allocation approaches for DSRC networks into those categories.

\begin{figure}[htbp]
  \centering
  \includegraphics[width=\figwidth]{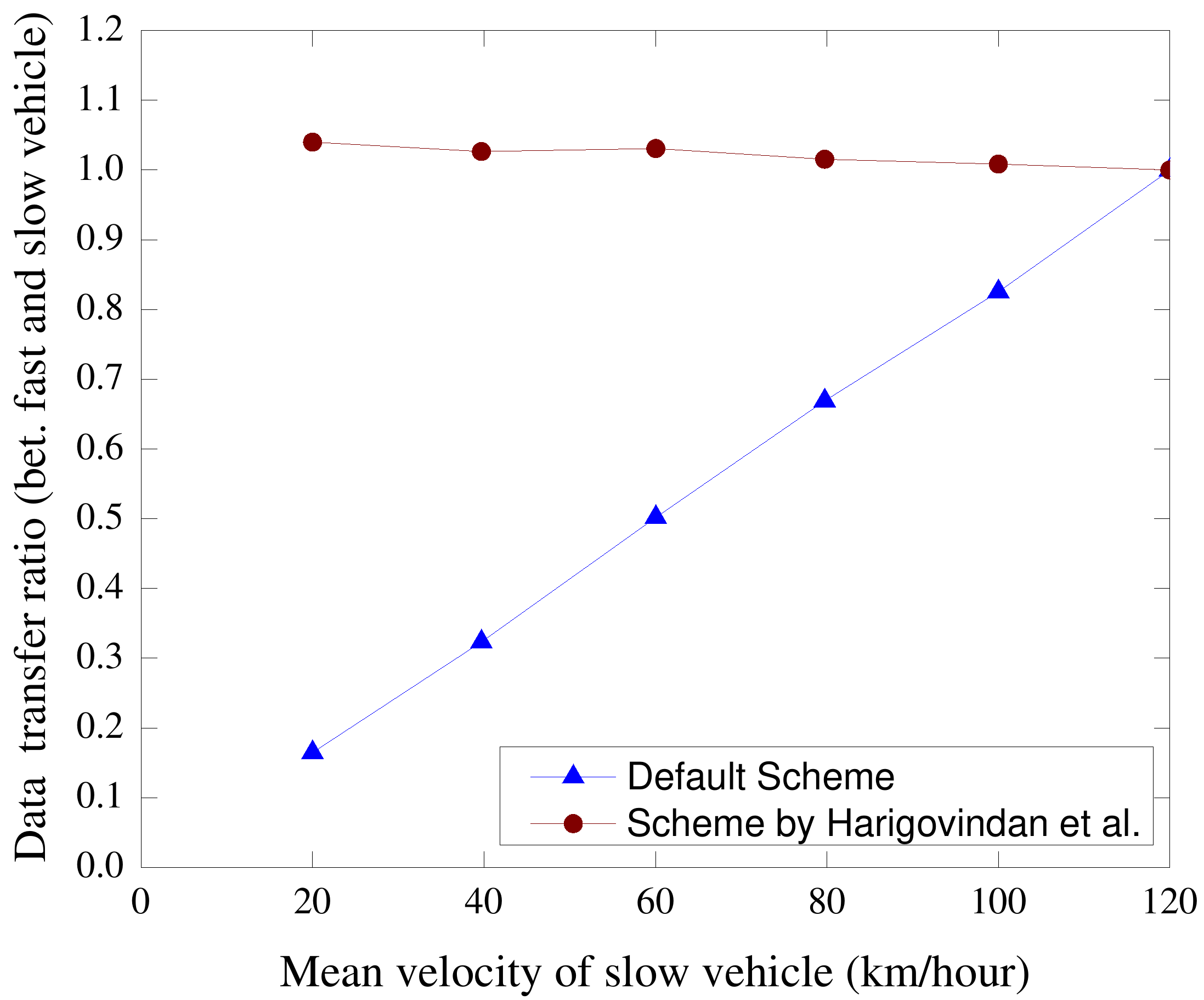}\\
  \caption{Data transfer ratio for fast and slow vehicles versus mean velocity of the slow vehicles. Comparison between default  DSRC  and  the  scheme proposed by Harigovindan et al. \cite{Harigovindan2012}.}\label{fig:Harigovindan2012}
\end{figure}

\begin{figure*} [h]
 \centering
  \begin{subfigure}[t]{0.5\textwidth}
 \includegraphics[width=\textwidth]{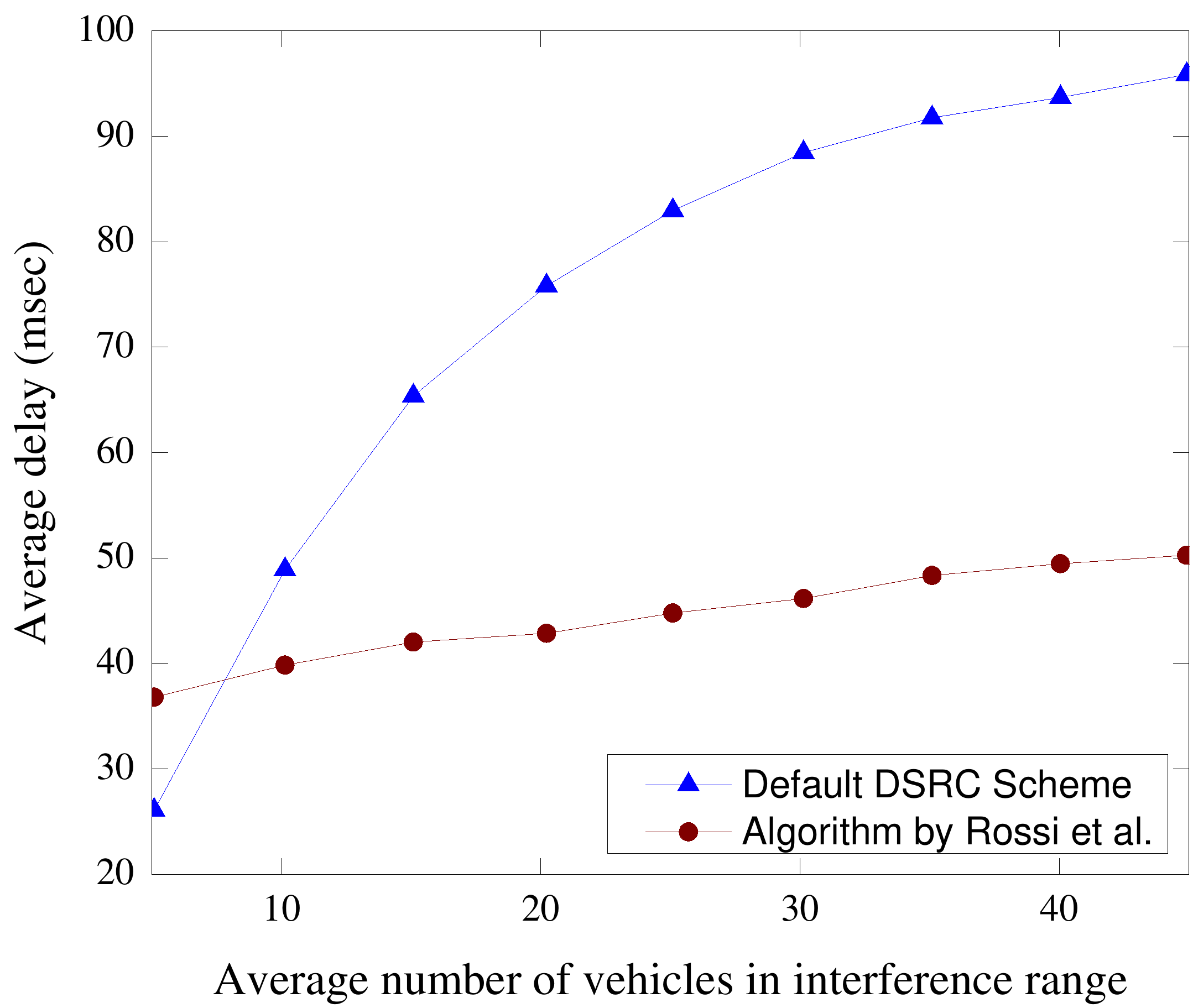}
  \caption{Average transmission delay.}
 \label{fig:Rossi2017a}
 \end{subfigure}~
 \begin{subfigure}[t]{0.5\textwidth}
\includegraphics[width=\textwidth]{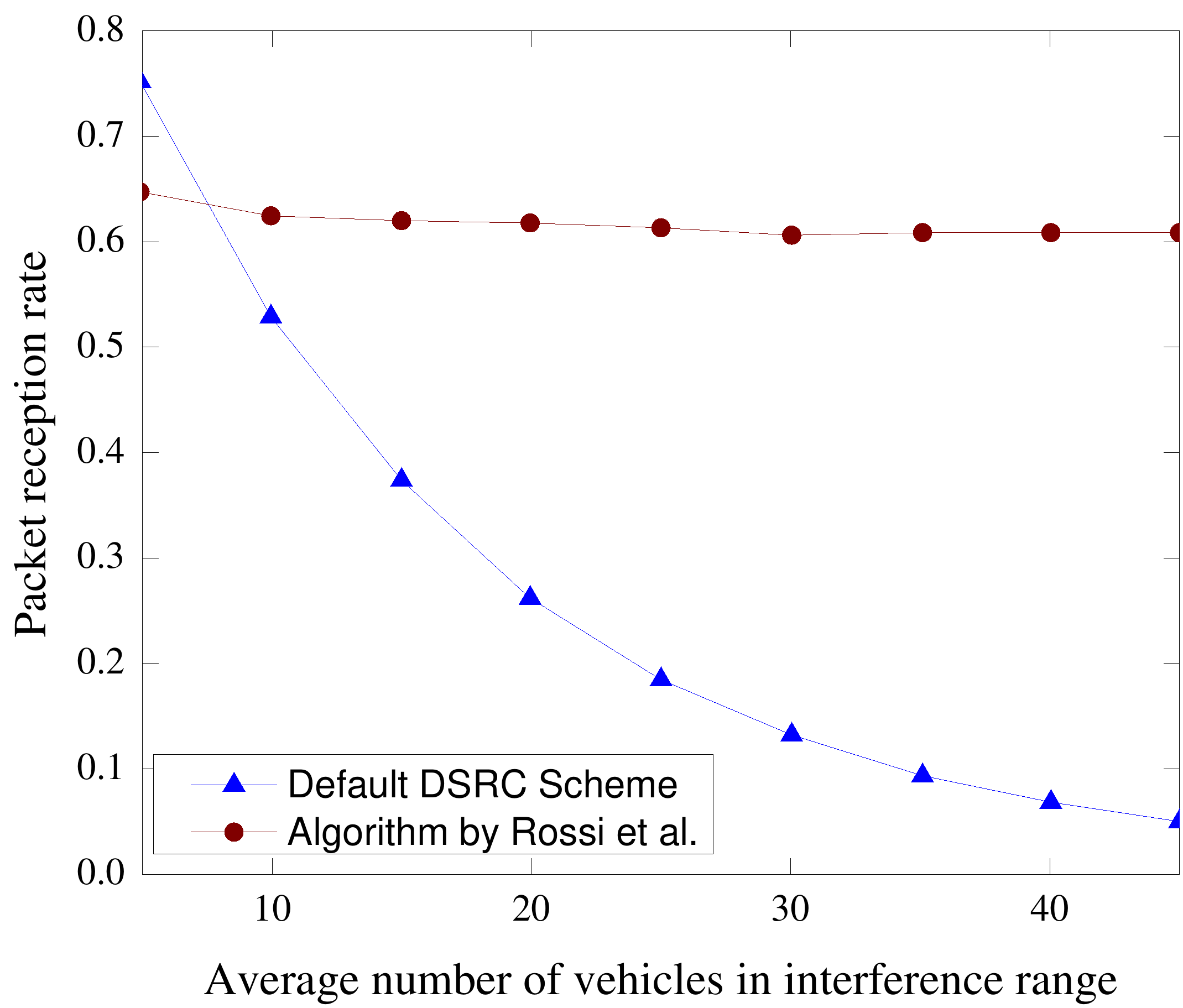}
\caption{Packet reception rate by adjacent vehicle.}
\label{fig:Rossi2017b}
 \end{subfigure}%
 \caption{Transmission performance with the stochastic model and algorithm proposed by Rossi et al. \cite{Rossi2017}.}\label{fig:Rossi2017}
\end{figure*}

\subsection{MAC Parameter Allocation}
In  a traditional DSRC network, all vehicles adopt identical MAC parameters by default and hence have equal opportunity to access the network resources. However, this setting may be unfair for fast moving vehicles compared to slow moving vehicles, potentially leading to significant degradation in network performance. For example, the throughput of a high speed vehicle may degrade significantly compared to a slow moving vehicle as the latter has a better chance to communicate with its RSU, due to its long residence time in the coverage area of the RSU.  Several studies have been carried out on MAC parameter allocation in DSRC networks to enhance reliability, throughput, and fairness.  Harigovindan et al. \cite{Harigovindan2012} presented a contention window allocation strategy to resolve the aforementioned unfairness problem and to dynamically adapt the MAC parameters based on the residence time of vehicles. Specifically, an optimal selection on the minimum contention window (required for any vehicle) has been derived by taking into consideration the mean speed of vehicles in the network. To validate the proposed technique, authors in \cite{Harigovindan2012}   simulated a V2I network  using an event driven custom simulation program (written in C++ programming language), where the MAC layer was based on the EDCA mode of IEEE 802.11p   and the physical layer was based on IEEE 802.11a. The mean velocity of the slow vehicle was set to  60 km/hr whereas the mean velocity of fast vehicle was set to 120km/hr. Fig. \ref{fig:Harigovindan2012} compares the default DSRC scheme  with the approach proposed in \cite{Harigovindan2012} in terms of the data  transfer ratio (for fast and slow vehicles)  versus  mean velocity of slow vehicles. It is observed that for the default DSRC  scheme, the data transfer ratio increases as the mean velocity of slow vehicles increases. In fact, in this case, the residence time of slowly moving vehicles decreases within a RSU's coverage area and hence the data transfer decreases correspondingly. On the other hand, a relatively flat data transfer ratio is maintained with Harigovindan et al. proposed contention window allocation scheme which ensures equal chances of communication with the RSU for both slow and fast vehicles\footnote{A contention window allocation approach similar to that in \cite{Harigovindan2012} can be found in \cite{Karamad2008}.}  Note that the proposed technique can cause unfairness to  slower vehicles in the event of a small number of fast vehicles and a large number of slow vehicles, as slower vehicles as they could experience higher levels of loss. Also, the proposed technique will likely cause unfairness in a situation when a highway lane is occupied with a platoon of slow moving vehicle, while an adjacent lane is occupied with a steady stream of faster vehicles.

To maximize throughput among neighboring vehicles, a stochastic model was proposed by Rossi et al.  \cite{Rossi2015,Rossi2017} to find the optimal maximum contention window using the surrounding  vehicle density.  By exploiting the equivalence between the slotted Aloha and the broadcast CSMA/CA protocols, an  amended  CSMA/CA  protocol was integrated   in the stochastic model   to  maximise the single-hop throughput among adjacent vehicles. To  validate  the  proposed model,  authors in  \cite{Rossi2017}  simulated (in Network Simulator 2 (NS-2))  a vehicular network  considering a  one-lane,  single-direction  road of length 5 km. In the simulation, it  is  assumed  that  vehicles  are  able  to  estimate the  number  of  neighbouring  vehicles  in  the  interference range.  The transmission range is set to be  100m, while setting  the  path  loss  exponent to 4.  Fig.~\ref{fig:Rossi2017}  shows that the proposed protocol in \cite{Rossi2015,Rossi2017} offers much lower average transmission delay as well as significantly improved packet reception rate (compared to the standard DSRC protocols) due to reduced packet collision with optimized contention window size.

\begin{figure*} [h]
 \centering
  \begin{subfigure}[t]{0.5\textwidth}.
 \includegraphics[width=\textwidth]{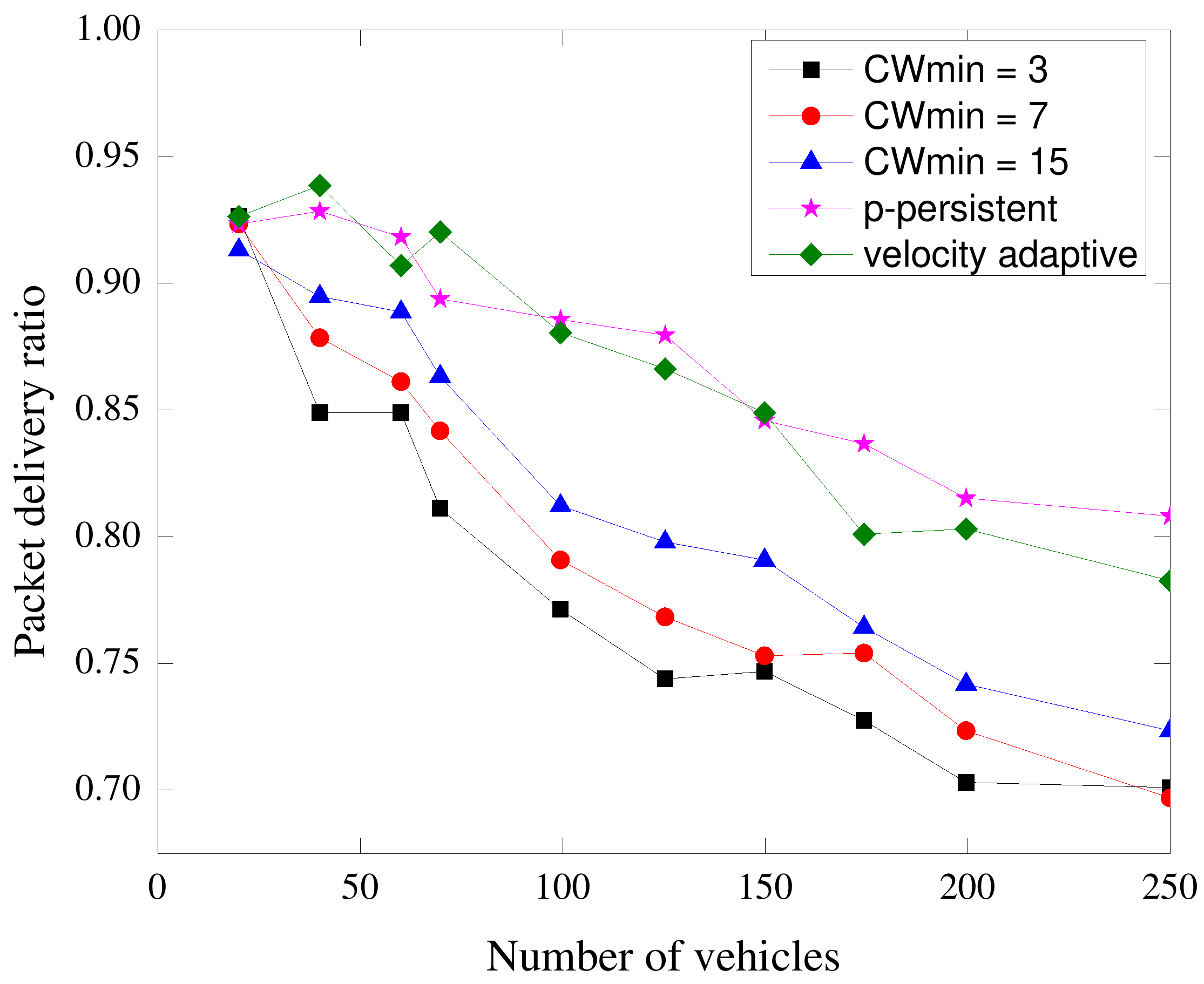}
  \caption{Packet delivery ratio.}
 \label{fig:Alasmary2012a}
 \end{subfigure}~
 \begin{subfigure}[t]{0.5\textwidth}
\includegraphics[width=\textwidth]{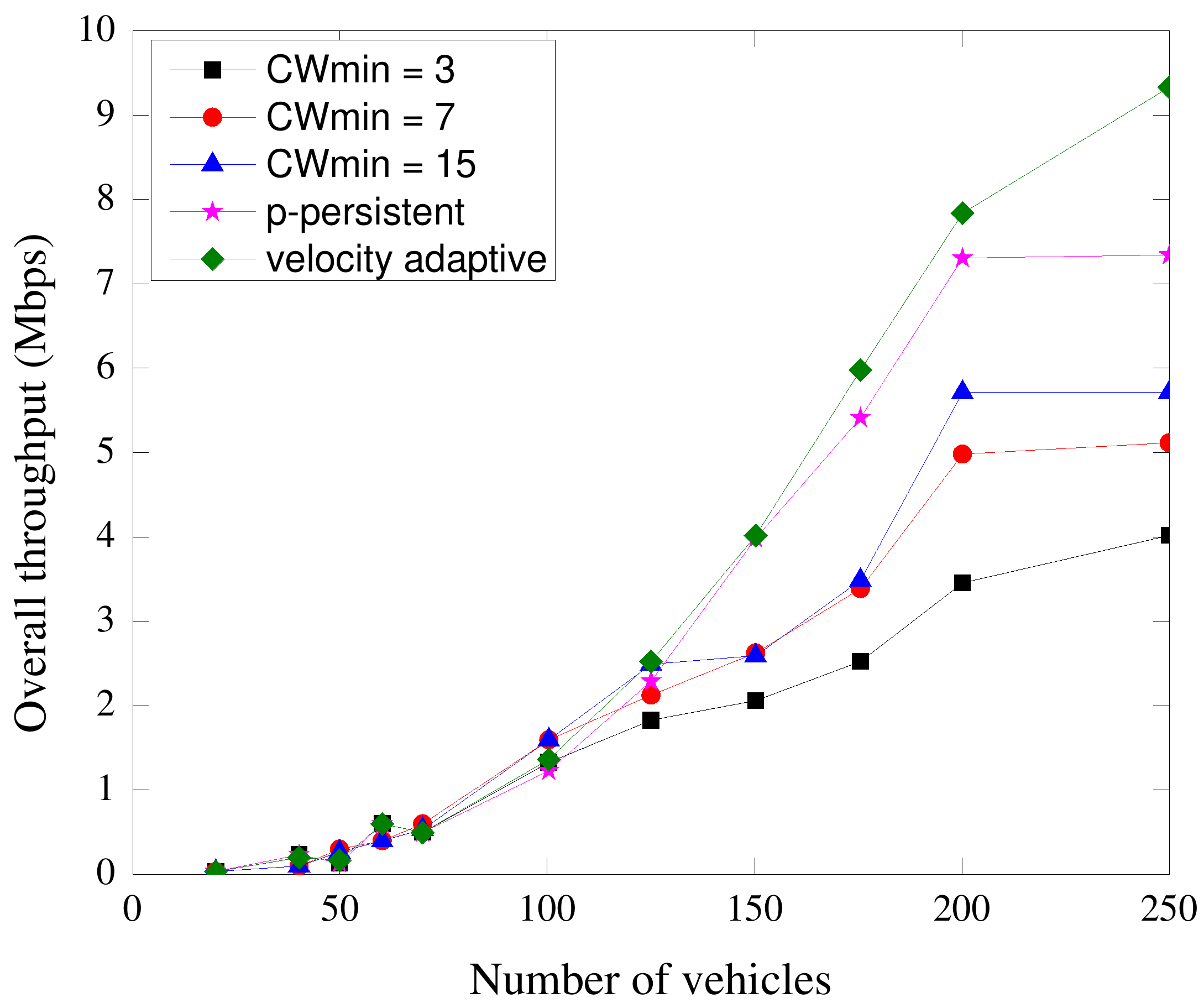}
\caption{Delay time.}
\label{fig:Alasmary2012b}
 \end{subfigure}%
 \caption{Network throughput simulation results   \cite{Alasmary2012} for different minimum Contention Window (CWmin) sizes.}\label{fig:Alasmary2012}
\end{figure*}

In  \cite{Alasmary2012},  two dynamic Contention Window (CW) allocation schemes are proposed to improve the network performance in high mobility environments. The first scheme is a p-persistent based approach \cite{Cali2000} which  dynamically assigns the contention window based on the number of neighbor vehicles, while the second scheme performs contention window adaptation based on  other vehicle's relative velocity. To evaluate the impact of the proposed dynamic allocation schemes, authors in \cite{Alasmary2012} simulated a Network Simulator 2 (NS-2) based a vehicular network considering   a 3-lane highway with a length of 5 km and a width of 10 m per lane. Same 802.11p MAC parameters were set  for all vehicles and   vehicles' velocities were varied from 60 km/h to 120 km/h.  Fig.~\ref{fig:Alasmary2012} compares their proposed schemes in terms of the packet delivery ratios and network throughput. It is observed that both schemes provide enhanced performance (compared to the default DSRC scheme with minimum contention window sizes  CWmin = 3,7,15) as they give rise to reduced packet collisions. Moreover, each scheme provides enhanced performance for a specific scenario. For example, the first scheme exhibits better packet delivery ratio when the number of vehicles in the network is large. In terms of network throughput, the second scheme outperforms the  first when the number of vehicles is higher than 80.

\subsection{Channel Allocation  for Emergency Messages}
DSRC/WAVE uses  orthogonal  frequency bands to  support   multi-channel  operation while considering equal share of available channels to all messages.   Emergency  messages (e.g., mission critical messages that carry safety-related information) in vehicular networks need to be  processed  with  high  priority, ultra reliability, and low latency. Ryu et al. \cite{Ryu2011}  proposed a   multi-channel   allocation strategy called DSRC-based   Multi-channel   Allocation   for   Emergency   message   dissemination (DMAE) by first identifying  the  available  bandwidth  of  channels  and then  allocating  the  channel with the largest  bandwidth to  the  emergency  message while maintaining QoS between RSU and OBU through periodic channel switching. It is shown that the emergency PDR of DMAE is higher than the PDR of WAVE as DMAE assigns   available  SCH  with  maximum  bandwidth  to the emergency  messages. Moreover,  DMAE  outperforms WAVE  in terms of delay performance as it  can  assign  emergency  messages  to  reserved  channels  in the event of heavy traffic scenario.

\begin{table*}
\caption{Different Modulation and Coding Schemes (MCS) and their corresponding data rates adopted in DSRC. BPSK: Binary Phase Shift Keying; QPSK: Quadrature Phase Shift Keying; QAM: Quadrature Amplitude Modulation.}  \label{Table:MCS}    
\scriptsize
\centering          
\begin{tabular}{|c|c|c|c|c|c|}    
\hline 
\textbf{MCS Index} & \textbf{Modulation} &  \textbf{Code rate} & \textbf{Data rate (Mbps)} & \textbf{Effective Data rate (Mbps)} & \textbf{Communication range (m)}  \\[0.5ex]  
\hline \hline
1& BPSK &$\frac{1}{2}$ & 3 & 2.77 & 1000 \\ \hline
2& BPSK &$\frac{3}{4}$ & 4.5 & 4.05  & 900  \\ \hline
3& QPSK &$\frac{1}{2}$ & 6 & 5.28 & 800 \\ \hline
4& QPSK &$\frac{3}{4}$ &  9 & 7.59 & 700 \\ \hline
5& 16-QAM &$\frac{1}{2}$ & 12 & 9.69 & 600  \\ \hline
6& 16-QAM &$\frac{3}{4}$ &18 & 13.59 &500 \\ \hline
7& 64-QAM &$\frac{2}{3} $& 24 & 16.64 &400 \\ \hline
8& 64-QAM &$\frac{3}{4} $& 27 & 18.09 & 300  \\ \hline
\end{tabular}
\end{table*}

\begin{figure*} [h]
 \centering
  \begin{subfigure}[t]{0.5\textwidth}
 \includegraphics[width=\textwidth]{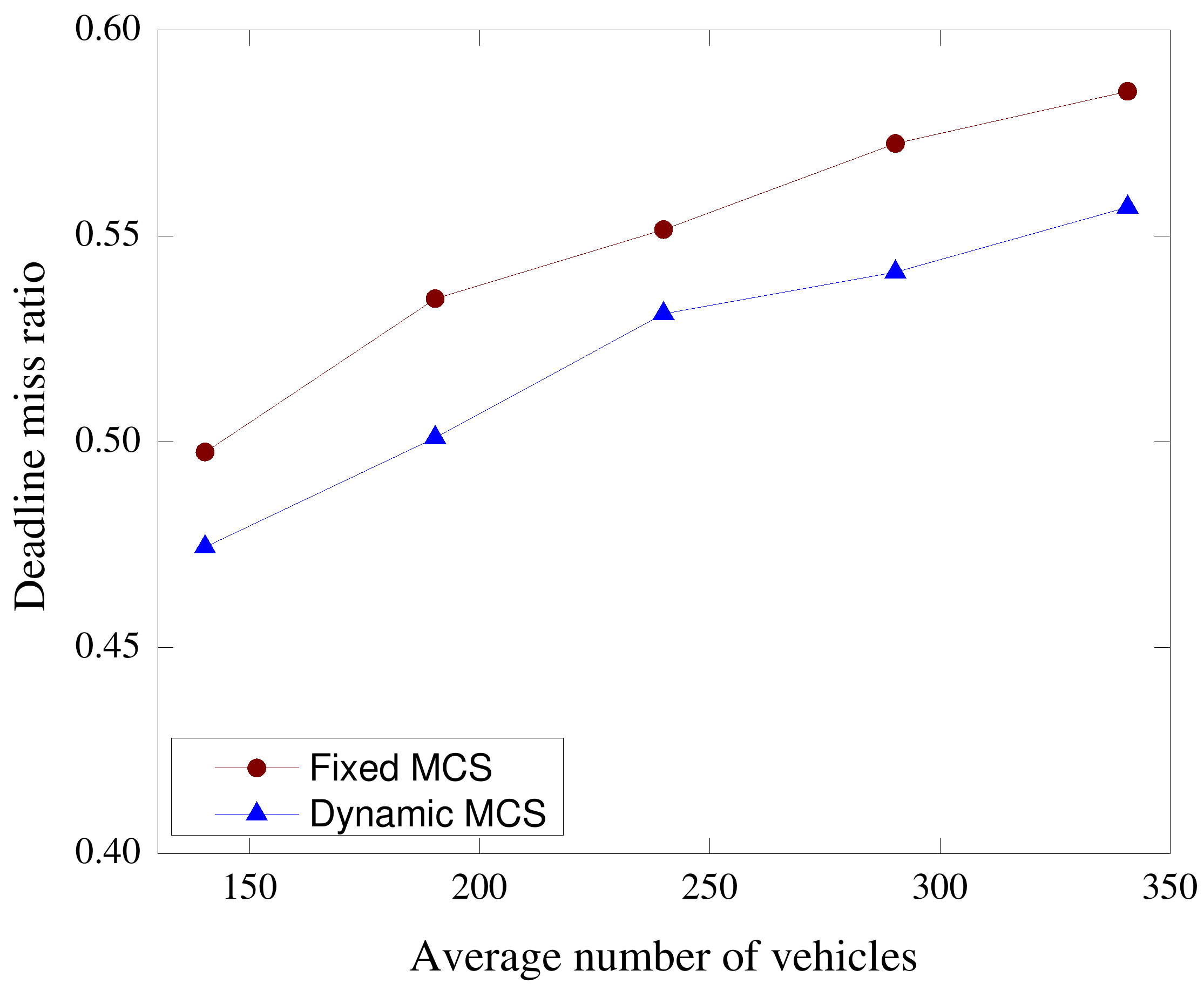}
  \caption{Deadline miss ratio.}
 \label{fig:Ryu2011a}
 \end{subfigure}~
 \begin{subfigure}[t]{0.5\textwidth}
\includegraphics[width=\textwidth]{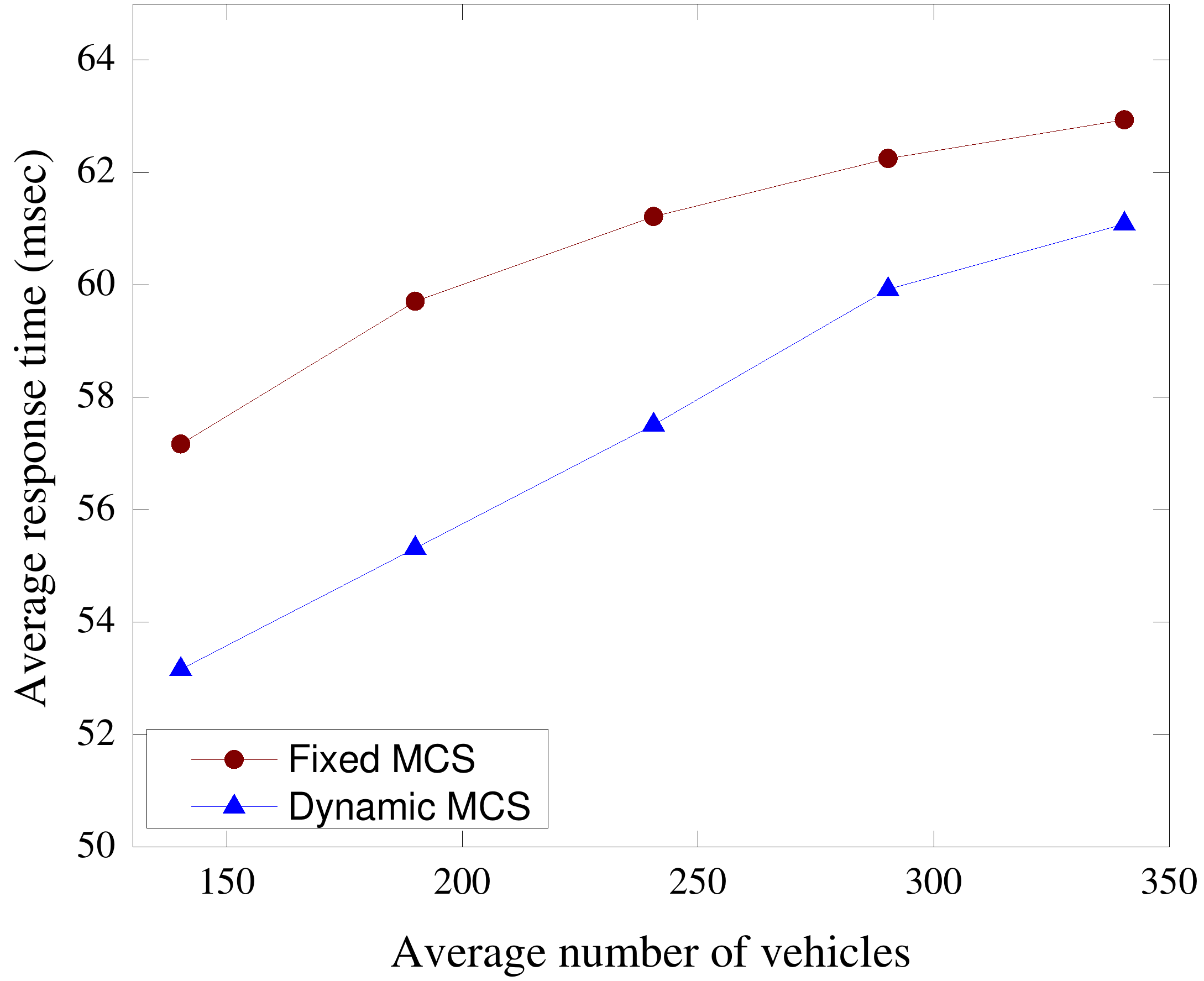}
\caption{Average response time.}
\label{fig:Ryu2011b}
 \end{subfigure}%
 \caption{Performance comparison between fixed and dynamic Modulation and Coding Schemes (MCS)  \cite{Ali2018}.}\label{fig:Ali_2018}
\end{figure*}

\subsection{Rate Allocation}
IEEE 802.11p based communication supports multiple MCS to allow a wide range of data transmission rates ranging from 3 Mbps to 27 Mbps. The  data rates (both nominal and average effective data rates  \cite{8422309}) and transmission ranges for different MCS are shown in Table \ref{Table:MCS}. For the sake of simplicity,  a constant MCS is often assumed in previous works on  vehicular communications. This strategy may  deteriorate the communication performance as constant MCS may not be suitable for diverse traffic environments in different roadway scenarios. More precisely, the IEEE 802.11 MAC protocol  offers  equal   transmission  opportunities to  the  competing  nodes  when  all  nodes  experience  similar channel  conditions.  However, with  varying  channel condition  and congested network,  throughput-based fairness will lead to drastically reduced aggregate throughput. As a solution, \cite{Sheu2010} proposed a new Vehicular Channel Access Scheme (VCAS) to maintain a trade-off between overall throughput and  fairness. In this scheme,  a  number  of  vehicles  with  similar  transmission  rates are grouped into  one  channel to  achieve  the  overall  throughput requirement,  while the   fairness\footnote{In the context of throughput of each vehicle.}  requirement is achieved by controlling the group  sizes. Grouping of OBUs with similar transmission rates boost the system throughput by eliminating  the  performance anomaly phenomena resulted from multiple transmission rates in the IEEE 802.11p multi-channel networks.   By adopting a  marginal  utility  model to allocate an appropriate  transmission rate per SCH (determined by  predefined transmission distance  thresholds), it is shown in \cite{Sheu2010} that their proposed scheme can simultaneously achieve enhanced fairness  and  overall system  throughput over the existing  scheme  adopted  in DSRC system. More recently,  \cite{Ali2018} proposed the allocation of variable MCS  (i.e., variable data rates) in network coding-assisted heterogeneous on-demand  data access, in which the MCS for disseminating data items were assigned based on the distance of the requested vehicles from the RSU.   Authors  devised a dynamic threshold based network coding for minimizing the system response time, where the   coded packet is  formed in such a way that  the coded packet always contains the most urgent request and the transmission time of the  coded packet does not exceed the deadline of the most urgent request. Note that the transmission time  depends on the size of the coded packet \footnote{Note that the size of the encoded packet is the size of the maximum size data items among all the data items that are being encoded in the coded packet.} and the selected  MCS that offers  highest data rate while ensuring serving of all the requests included in the coded packet. We  evaluated the performance of the proposed scheme  by simulating an urban grid-type multi RSU vehicular network. We have  implemented the simulation model  using CSIM19 \cite{977280} and conducted  the simulation  using the default settings of IEEE 802.11p PHY and MAC layer standard.  The vehicle's mobility was modelled by following a  Manhattan  mobility  model.   Performance of the proposed scheme was evaluated in terms of deadline miss ratio and response time. Fig.~\ref{fig:Ali_2018} shows the performance comparison between fixed MCS  and dynamic MCS schemes.     Simulation results show that dynamic MCS scheme is capable of improving the on-demand requests serving capability and reducing the system response time.

\begin{table*}[h]
\centering

\caption{Existing RA techniques for C-V2X vehicular network.}  \label{tab:comp_cv2x}
\scriptsize
\centering          
\begin{tabular}{| p{1cm} | p{1.7cm}| p{1.5cm}| p{2.5cm} |p{2.5cm}|p{2cm}|p{1.2cm}|c|c|c|c|}    
\hline 
Reference  &   Scenario   & Use Case & Allocation Method & Allocation Objective & Allocated Parameters  & BS/RSU Assisted & Mobility \\[0.5ex]  
\hline

\cite{Zhang2013} & Single-lane Highway  & Generic & Graph theory & Maximizing throughput & Bandwidth   & \cmark &  \cmark  \\
\hline
\cite{Meng2018}  & Single-lane Highway  & Generic & Graph theory & Maximizing connectivity & Bandwidth   & \cmark &  \xmark  \\
\hline
 \cite{Liang2017}   & Multi-lane Highway  & Generic & Hungarian method & Maximizing ergodic capacity, reliability & Bandwidth, Power   & \cmark &  \cmark  \\
\hline
\cite{Sun2016}  & Urban grid layout  & Generic & Karush–Kuhn–Tucker
theory& Maximizing sum-rate; minimize latency & Bandwidth, power   & \cmark &  \xmark  \\
\hline
\cite{Zhang2016}   &  Urban grid layout; Single-lane Highway  & Generic & Perron–Frobenius theory & Maximizing concurrent reuses & Bandwidth   & \cmark &  \xmark  \\
\hline
\cite{Guo2019}    & Two-way urban roadway  & Generic & Hungarian method & Maximizing sum rate & Bandwidth, power   & \cmark &  \cmark  \\ 
\hline
  \cite{Lin2018}   & Multi-RSU network  & Fog computing & Lagrangian algorithm & Maximizing utility model  & Bandwidth   & \cmark &  \xmark  \\
  \hline
  \cite{Lin2018a}   & Multi-RSU network & Cloud computing &  Semi-Markov  decision   process & Maximizing  discount value & computing  resource   & \cmark &  \xmark  \\
\hline
 \cite{Ahmed2018}   & Urban area  & Security & Dynamic semi-persistent method & Maximizing resource utilization  & Bandwidth   & \cmark &  \cmark  \\
\hline
\cite{Yang2017}  &  Highway  &  Security & Greedy  algorithm & Maximizing secrecy rate & Bandwidth   & \cmark &  \cmark  \\
\hline
\cite{Wang2018} & Single-lane Highway  & Vehicle Platooning & Weight matching theory & Maximizing sum rate & Bandwidth   & \cmark &  \cmark  \\
\hline
\cite{Meng2018a}    & Single-lane Highway  & Vehicle Platooning & Lyapunov optimization & Maximizing service-guaranteed users & Bandwidth   & \cmark &  \cmark  \\
\hline
\cite{Gonzalez2019}  &  Highway  & Vehicle Platooning & Conflict-Free SPS & Maximizing stability & Bandwidth   & \cmark &  \cmark  \\
\hline
\cite{Scheuvens2019}  & Highway  &  Automated guided  vehicle & Application-adaptive  algorithm  & Maximizing QoS & Bandwidth   & \cmark &  \cmark  \\
\hline
\cite{Hung2016}  & Highway  & Vehicle Platooning & Lyapunov optimization & Minimizing delay, re-allocation rate & Bandwidth   & \cmark &  \cmark  \\
\hline
\cite{Peng2017}  & Highway  & Vehicle multi-platooning & Lyapunov optimization & Minimizing delay, transmission power & Bandwidth, power   & \cmark &  \cmark  \\
\hline
 \cite{Yang2016}   & Urban grid layout  & Generic   & Subpool sensing-based algorithm  & Minimizing interference & Bandwidth   & \xmark &  \cmark  \\
\hline
\cite{Sahin2018}    & Single-lane Highway  & Generic & Pre-scheduling & Maximizing reliability & Bandwidth   & \xmark &  \cmark  \\
\hline
 \cite{our_TVT_2019}  & Intersection  & BSM relaying & Exhaustive search algorithm & Minimizing interference & Bandwidth   & \xmark &  \xmark  \\
\hline
\end{tabular}

\end{table*}

\section{Resource Allocation in C-V2X}
The capability of supporting diverse vertical applications and use cases is a major feature of 5G communication systems and beyond. Examples of vertical use cases include smart homes/cites, e-health, factories of the future, intelligent refineries and chemical plants, and Cellular V2X (C-V2X). A strong catalyst for deeper and wider integration of wireless communications into our lives, C-V2X has been advocated by many mobile operators under the evolution of 3GPP's LTE and 5G NR \cite{Huawei2017}. Compared to DSRC, C-V2X acts as a ``long-range sensor" (aided by sophisticated cameras, radar, lidar, RSUs, cellular infrastructure and network) to allow vehicles to see/predict various traffic situations, road conditions, and emergent hazards several miles away.

From a network point of view, there are three major 5G use cases to be supported: enhanced Mobile Broadband (eMBB) communications, massive Machine-Type Communications (mMTC), Ultra-Reliable and Low-Latency Communications (URLLC). As far as C-V2X is concerned, eMBB, aiming to provide data rates of at least 10 Gbps for the uplink and 20 Gbps for the downlink channels, plays a pivotal role for in-car video conferencing/gaming, various multimedia services, or high-precision map downloading, etc; mMTC will allow future driverless vehicles to constantly sense and learn the instantaneous driving environments using massive number of connected sensors deployed in-car or attached to the infrastructure; URLLC, targeting to achieve 1 ms over-the-air round-trip time for a single transmission with reliability of at least $99.999\%$ will be instrumental for example for autonomous emergency braking and hazard prevention.

However, C-V2X has to share and compete with other vertical applications for system resources (e.g., spectrum/network bandwidth, storage and computing, etc) under a common physical infrastructure. RA for C-V2X therefore is a trade-off with a variety of data requirements from different vertical applications. A central question is how to design an efficient network to provide guaranteed quality of service (QoS) for C-V2X while balancing the data services for other vertical applications.

\subsection{RA for Traditional Cellular Systems}


\begin{figure}[htbp]
  \centering
  \includegraphics[width=\figwidth]{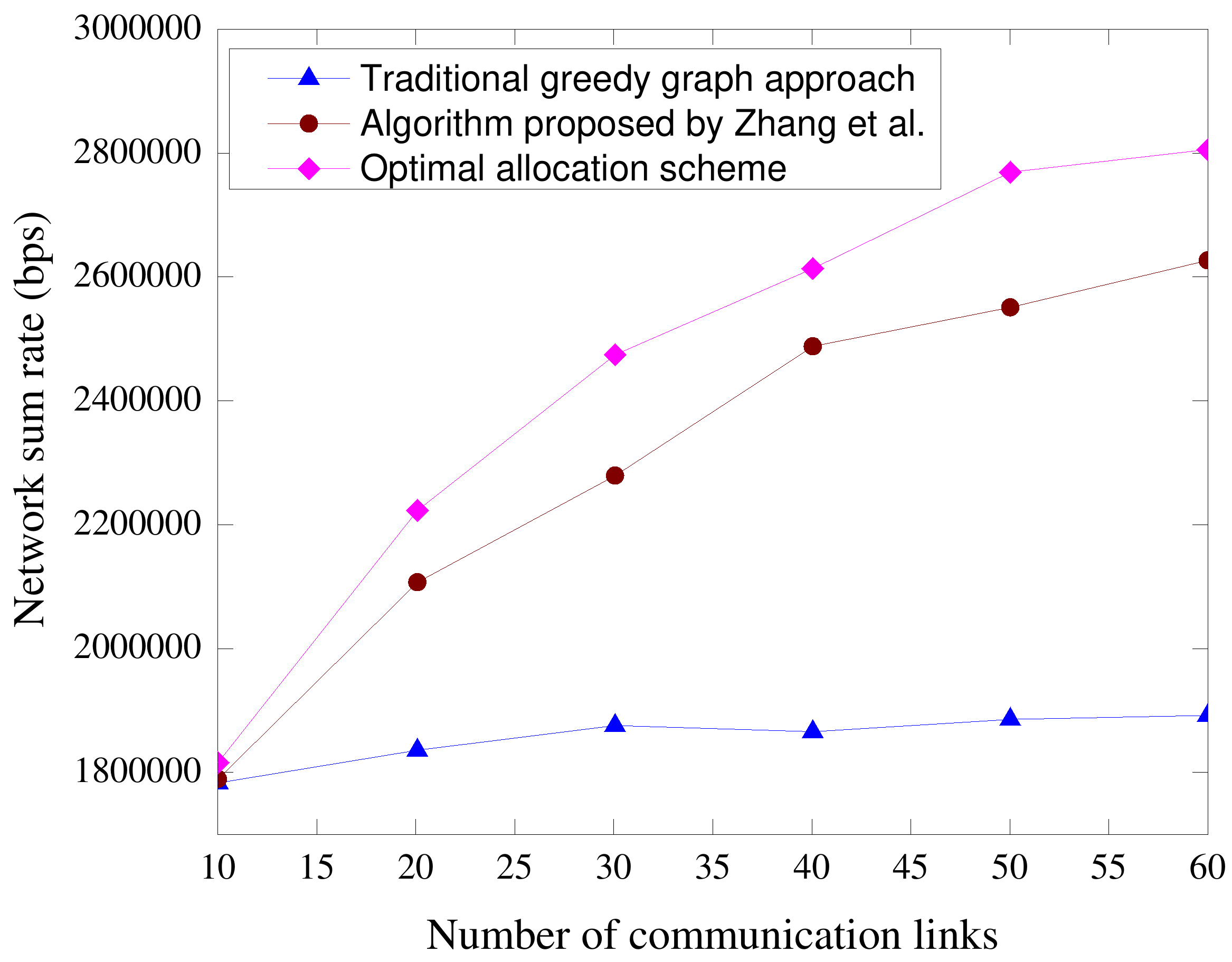}\\
  \caption{Sum-rate comparison between traditional scheme and optimal scheme proposed by Zhang et al. \cite{Zhang2013}.}\label{fig:Zhang2013}
\end{figure}

\begin{figure*} [h]
 \centering
  \begin{subfigure}[t]{0.5\textwidth}
 \includegraphics[width=\textwidth]{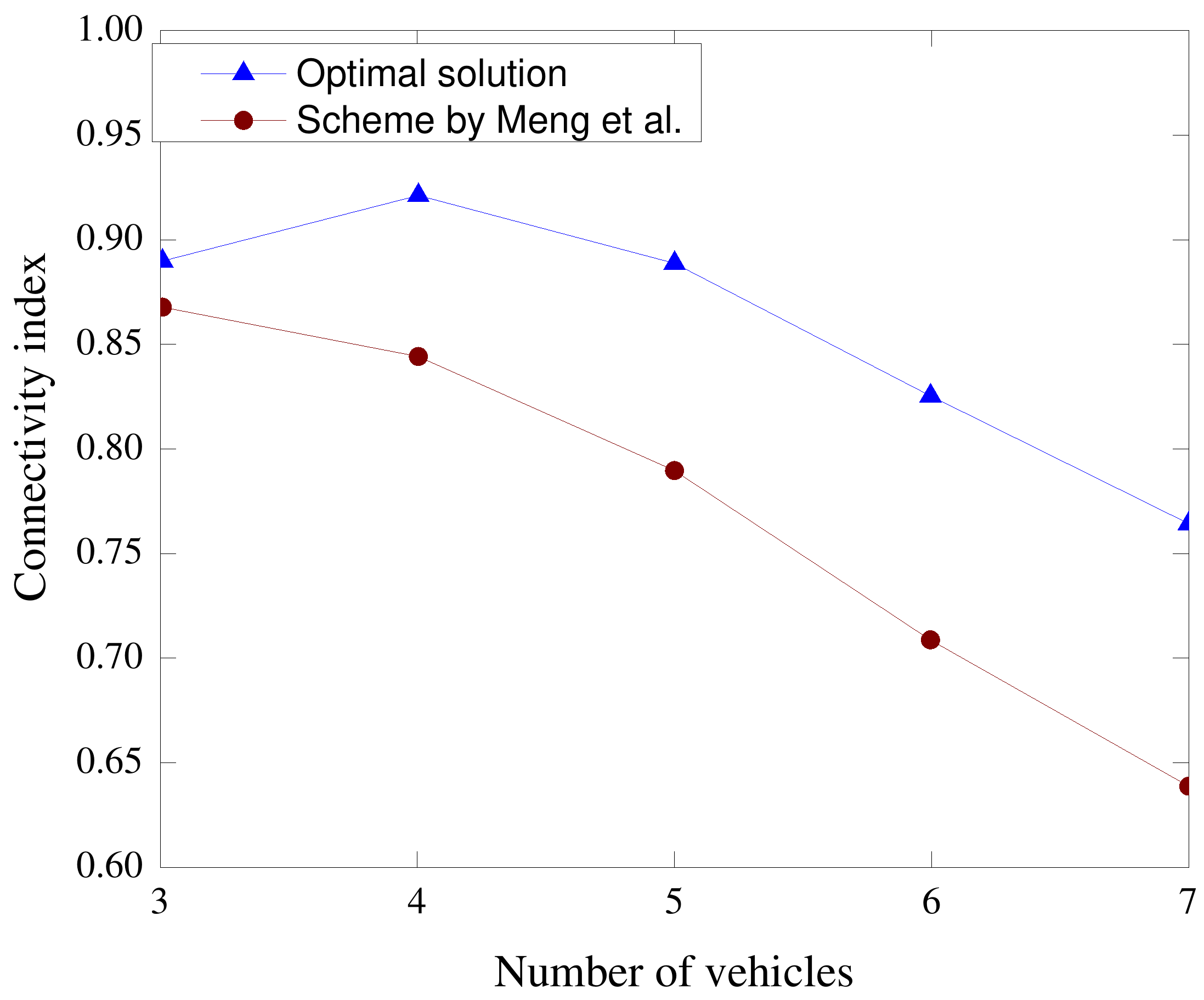}
  \caption{The gap between the optimal solution and the proposed sub-optimal solution.}
 \label{fig:Meng2018a}
 \end{subfigure}~
 \begin{subfigure}[t]{0.5\textwidth}
\includegraphics[width=\textwidth]{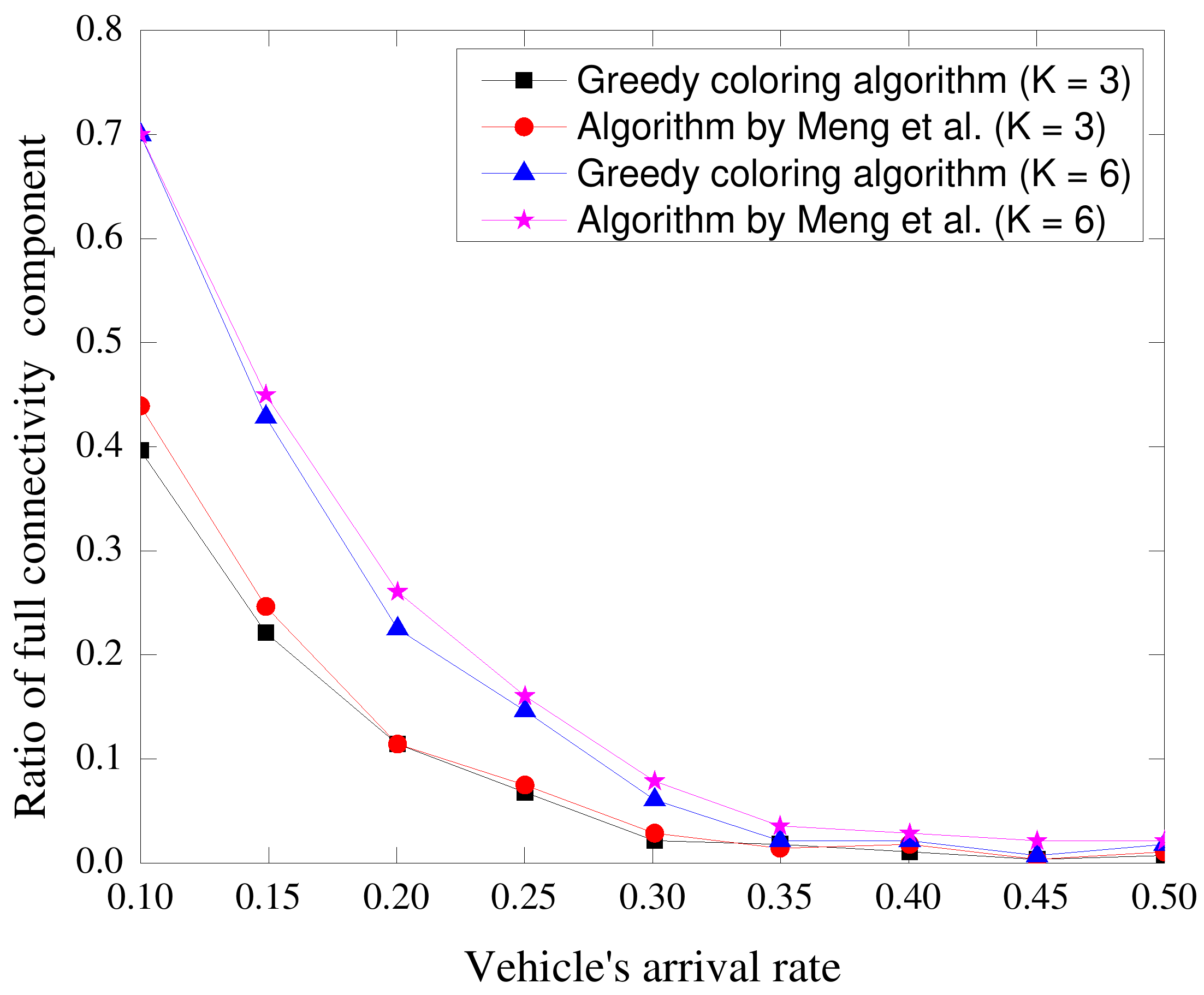}
\caption{The percent of the full connectivity (FC) components for different vehicle arrival rates.}
\label{fig:Meng2018b}
 \end{subfigure}%
 \caption{Performance of the RA scheme proposed by Meng et al. \cite{Meng2018}.}\label{fig:Meng2018}
\end{figure*}

Graph based interference aware RA strategies have been proposed in \cite{Zhang2013,Meng2018}, where the weights of the edges are assigned according to the interference terms between the related vertices. The scheme proposed by Zhang et al. \cite{Zhang2013} formulates an optimization problem with the objective of maximizing the network sum rate \footnote{Network sum rate  is defined as the sum of the channel capacity for all V2I and V2V communication links within the network.} with low computational complexity.  Considering the interference between different communication links, authors formulated the resource-sharing problem as a resource assignment optimization problem for a vehicular network scenario,  where different V2V and V2I communication links are permitted to access the same resources for their individual data transmission. To avoid high computational complexity, graph theory was used to effectively obtain a suboptimal resource assignment solution.  Authors in \cite{Zhang2013} conducted a simulation considering a $20 \text{m} \times 500 \text{m}$  road layout  with a base station located at the center of the long edge. The vehicles were distributed randomly  within the road with a random velocity of between $0–100$ km/h. The interference  radius of vehicle and base station were set to 10 m and 100 m, respectively. For resource allocation purpose, number of  resource blocks was set  to 10.   It is shown in Fig.~\ref{fig:Zhang2013} that their proposed scheme exhibits higher network sum rate than the traditional orthogonal communication mode. In contrast, the work in  \cite{Meng2018} aims at improving the connectivity of vehicular communications by introducing a metric called \textit{connectivity index}, which is obtained from the percentage of vehicles in the network being assigned with resources while satisfying the interference constraints. With the aid of the minimum spanning tree approach \cite{West1996}, Meng et al. \cite{Meng2018} proposed a RA algorithm to improve the connectivity of the network.   Authors in \cite{Meng2018} evaluated the performance of the proposed scheme using simulation (using NS-3) where  a two-way four-lane road of 1 km with randomly distributed vehicles was considered. The transmission radius of vehicles was assumed to be 50 m, while the speed of the vehicles varied from 20km/h to 60 km/h. Fig.~\ref{fig:Meng2018} shows the performance of the RA scheme proposed in \cite{Meng2018}. The connectivity index performance is presented in Fig.~\ref{fig:Meng2018a} with varying number of vehicles, whilst the  performance of a brute force search algorithm is shown as a benchmark. We observe that the connectivity index of Meng et al.'s algorithm is only 17.1\% away from the optimum solution obtained with the brute force search algorithm.  In Fig.~\ref{fig:Meng2018b}, we present the full connectivity performance of the algorithm proposed in \cite{Meng2018} and compare with a greedy graph coloring algorithm \cite{Etzion1998}. We observe a similar full connectivity performance for both algorithms, while the graph coloring algorithm exhibits high computational complexity.   As expected, the full connectivity percentage decays with the increase of  vehicle arrival rate (i.e., denser vehicular network).

By  exploiting  geographical information, \cite{Liang2017} proposed a joint RA and power control scheme for reliable D2D-enabled vehicular communications by considering slow fading channel information. Queuing dynamics was also considered in \cite{Liang2017} in order to meet the requirements of different QoS in vehicular networks. \cite{Sun2016} developed a heuristic algorithm, named Separate  resOurce  bLockand  powEr  allocatioN (SOLEN), under large-scale vehicular fading channels to maximize the sum rate of cellular users while satisfying the vehicular users' requirements on latency and reliability. Similar to \cite{Sun2016}, \cite{Mei2018} incorporated dynamic MCS in the process of RBs and transmit  power allocation  for guaranteed reliability and latency. It is shown that  by adopting dynamic MCS in the allocation algorithm, the algorithm proposed in \cite{Mei2018} outperforms that of \cite{Sun2016}  in terms of average  outage probability and packet latency. To support D2D-based safety-critical vehicular communication, a cluster-based RA scheme was proposed in  \cite{Sun2016b} by maximizing the cellular users' sum rate. This is achieved by a three-step heuristic algorithm with the knowledge of the slowly varying channel state information of uplink channel.

The work in \cite{Zhang2016} proposed a centralized  RA  algorithm by utilizing the spectral radius estimation theory.  Their proposed algorithm  maximizes the number of concurrent reuse of resources by multiple vehicles instead of maximizing the sum rate  (a method often used in traditional allocation algorithms). With eNodeB centrally deciding the resource reuse for the vehicles in the network, the scheme proposed in \cite{Zhang2016} exhibits  significant  improvement in  the spectrum efficiency and demonstrates the capability of maintaining the required QoS when the vehicle density is high. \cite{Guo2019} proposed a RA scheme to support V2X communications in a D2D-enabled cellular system, where the V2I communication is supported by a traditional cellular uplink strategy and the V2V communication is enabled by D2D communications in  reuse  mode. \cite{Guo2019} formulated  an optimization problem to  maximize the sum ergodic capacity of the vehicle-to-infrastructure (V2I) links while satisfying the delay requirements of  V2V links. The optimization problem was solved by  combining a  bipartite  matching algorithm and effective capacity theory.



\subsection{RA for Vehicular Computing Systems}
In recent years, integration of  vehicular network  with  mobile cloud  computing, also known as  vehicular  computing  system, has attracted increasing interest for its capability of providing  real-time  services  to  on-board users \cite{Gerla2012,Bitam2015}. RA for vehicular computing systems has been investigated in  \cite{Lin2018a,Lin2018}. In particular, \cite{Lin2018a} integrated the computational resources of vehicles and RSUs in the vehicular cloud computing system to provide optimum services. The integration was performed by establishing a semi-Markov decision model for resource allocation in the vehicular cloud computing system, which allocates either vehicular  cloud  (consisting  of  vehicle’  computing  resources)  or remote clouds to handle vehicles’ service requests. Besides cloud computing, which is a centralized system,  fog computing is an attractive option for vehicular computing as it allows distributed decentralized infrastructure.  \cite{Lin2018} aimed to reduce the serving time \footnote{The serving time is the time required to serve a specific request, while serving method refers to the specific way to serve the request.} by optimally allocating the available bandwidth in a vehicular fog computing system. The optimization problem of \cite{Lin2018}, formulated based on the requirements of the serving methods, was solved in the following two steps: 1) finding the sub-optimal solutions by applying  the Lagrangian algorithm; 2) performing selection process to obtain the optimum  solution.

\subsection{RA for Secure Vehicular Networks}\label{sec:RA_Security_CV2X}
RA may also be exploited to enhance the secrecy of cellular vehicular networks.  By observing that LTE-based  V2X communication cannot properly preserve the  privacy, \cite{Ahmed2018} evaluated the message  delivery with  specified  security. A joint channel and security key assignment policy  was presented in  \cite{Ahmed2018} to enable a robust and secure V2X message dissemination.   The proposed approach classified V2X messages into four  categories  and utilized V2X interfaces and resource allocation mode (dynamic/ semi-persistent) intelligently to protect privacy.  Specially for the emergency message, a novel  random access with status feedback based resource allocation strategy was proposed in sidelink PC5 interface  to protect the privacy. In \cite{Yang2017}, a  RA scheme was proposed to enhance the  physical layer security in  cellular vehicular communication. A max-min secrecy rate based problem was formulated to allocate power and sub-carrier while taking into account the outdated Channel State Information (CSI) due to the high mobility.  The problem was solved in two stages: (i) with fixed  sub-carrier assignment, allocating the power level by using  a  bisection  method  allocation  problem; (ii) finding suboptimal sub-carrier allocation  by using greedy  algorithm.

\subsection{RA for Vehicle Platooning}
In recent years, vehicle platooning networks have been gaining growing research interest as they can lead to significant road capacity increase. In  \cite{Wang2018}, the authors proposed a RA scheme for D2D based vehicle platooning  to share control information efficiently and timely.  A time-division based intra-platoon and minimum rate guaranteed inter-platoon RA scheme was proposed to allocate the resources within the platoon, while ensuring optimized cellular users' rate. Moreover, to obtain a stable platoon, a formation algorithm was proposed in \cite{Wang2018} based on a leader evaluation method. Authors in \cite{Meng2018a} presented a RA strategy to reduce the re-allocation rate that enhances the number of guaranteed services in a vehicle platooning network. A time  dynamic  optimization problem was formulated in \cite{Meng2018a} under the constraint  of a network re-allocation rate. To further reduce the computational complexity,  their proposed optimization problem was converted into a deterministic optimization problem using the Lyapunov optimization theory \cite{Georgiadis2006}. Joint optimization of communication  and  control in vehicle platooning was proposed in \cite{Mei2018a}. An improved platooning system model was developed by taking into account both control and communication factors in vehicle platooning. A safety message dissemination scenario was considered under an LTE based vehicular network, where the platoon leader vehicle coordinates the allocation of available communication and control resources. A joint optimization problem of RB allocation and control parameter assignment was formulated with the constraints of communication reliability and  platoon stability. Through simulation results, it was shown that their proposed RA algorithm reduces the tracking error while maintaining the stability of the platoon. For cooperative adaptive cruise control (CACC) enabled platooning, a semi-persistent scheduling approach for LTE-V2X network was studied in \cite{Gonzalez2019,Gonzalez2019a,Gonzalez2019b}. A theoretical framework was  developed to find the required  scheduling  period that fulfills the string stability condition for CACC.  The scheduling  framework took into account different control and communication parameters such as   platoon  kinematics,  number of radio blocks,  packet sizes.  To reduce the  average  amount  of  links  provisioned, \cite{Scheuvens2019} proposed an adaptive resource allocation approach for automated guided vehicles (AVGs), where a  control communication co-design scheme was considered.  Authors have   derived  co-design   recommendations  to  improve the  correct  operation  of  AVGs, while considering the impact of packet loss on the  system.  It is shown  that   the impact of packet loss is not as severe as  commonly  assumed   with  appropriate  system  design. A dynamic resource re-allocation  technique was proposed in \cite{Hung2016} for the vehicle platooning scenario to  reduce the re-allocation rate and guarantee the delay requirement for each vehicle.   The proposed allocation algorithm aims to minimize the  process cost which is defined as the cost of signaling to the network due to the execution  of  resource  re-allocation. A closed form of the  resource re-allocation rate and the delay upper  bound was derived using Lyapunov optimization.  In \cite{Peng2017}, a joint sub-channel allocation scheme and   power   control   mechanism   were   proposed   for   LTE-based inter-vehicle  communications  in  a  multi-platooning  scenario. Authors performed  intra-  and  inter-platoon  communications  by combining  the  evolved  multimedia  broadcast  multicast  services (eMBMS) and device-to-device (D2D) multicast communications while ensuring  a desired trade-off between the required cellular resources and minimum  delay requirement.

\begin{figure} [htbp]
 \centering
  \includegraphics[width=\figwidth]{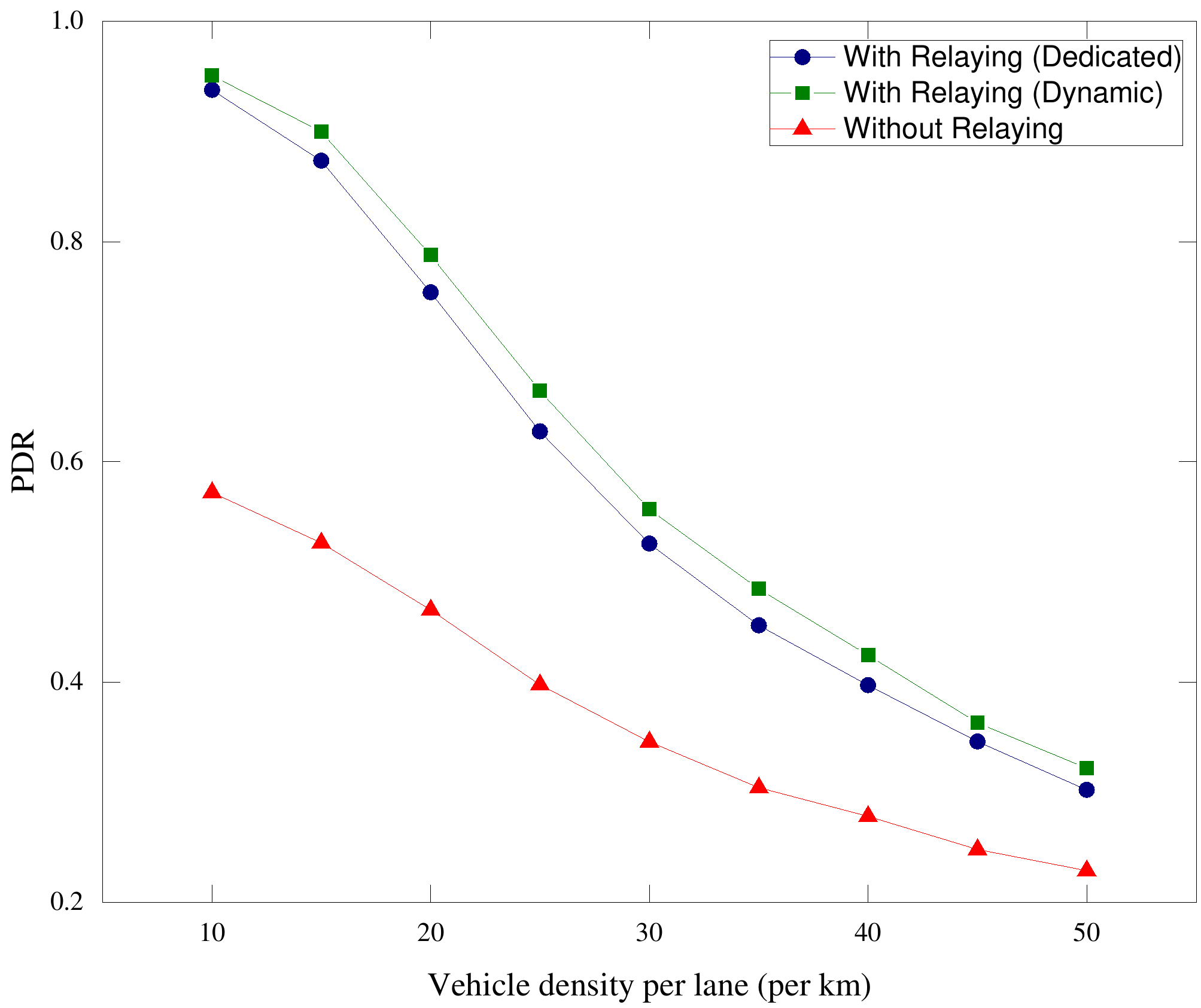}\\
 \caption{Packet delivery ratio (PDR) performance comparison when tagged is located with 100m from the intersection center.}\label{fig:our_TVT_2019}
\end{figure}

\subsection{RA for Out-of-Coverage Scenario}
A two-step distributed RA scheme was proposed in \cite{Yang2016} for out-of-coverage (i.e., out of eNodeB coverage) LTE V2V communication. In the first step, RBs are assigned based on the heading directions of vehicles. In other words, the same set of RBs are assigned to  the vehicles  moving in the same direction. In the second step, a channel sensing based strategy is utilized to avoid the packet collision between the vehicles which travel in parallel on the road. Recently, authors in \cite{Sahin2018} studied RA scheme for a delimited out-of-coverage scenario,  where the network infrastructure assigns the resources to vehicles based on the estimated location of vehicles.  The network infrastructure performs the resource allocation  based on the propagation conditions and the predictions of vehicle locations inside the out-of-coverage area. The past locations of the vehicles  are  used by the network infrastructure  to predict future trajectories of the vehicles and to predict the dwelling time of the vehicles inside the  out-of-coverage area.  The performance of  the proposed resource allocation scheme was analysed for non-scheduled services as well as  pre-scheduled services. More recently, authors in \cite{our_TVT_2019} analyzed and evaluated the safety message broadcasting performance of LTE-V2V out-of-coverage mode in an urban intersection scenario. In the context of vehicle assisted relaying,   two resource allocation strategies were presented, namely relaying with dedicated resources and relaying with dynamic resources. With the first strategy, resource blocks were reserved for the relaying vehicle, while for the latter strategy, the relaying vehicle dynamically finds the candidate resource  blocks with least interference.    To evaluate the performance, we have performed simulations modeling a 2km $\times$ 2km road network where the intersection-center is assumed at the middle of the road network. This simulation model was implemented using the LTEV2Vsim simulator presented in \cite{Cecchini2017}, where  the LTEV2VSim was extended by adding the intersection topology. The simulation scenario assumed  three lanes per travel direction with uniformly distributed (generated in random locations)  vehicles along the road.  The vehicular mobility was modelled by assigning an average speed of 50.08 km/h with a 3.21 km/h standard deviation.   Fig.~\ref{fig:our_TVT_2019} shows the performance of the proposed schemes when the transmission/target is located with 100m from the intersection center.  We observe that the relaying with dynamic resources gives slightly better performance than the relaying with dedicated resources. We also observe that the proposed relaying schemes exhibit significant broadcast performance improvement over the scheme without relaying when the vehicle density is low to moderate.

\begin{table*}[h]
\centering    

\caption{Existing RA techniques for heterogeneous vehicular networks.}  \label{tab:comp_het}
\scriptsize
\centering          
\begin{tabular}{| p{1cm} | p{1.5cm}| p{1.5cm}| p{1.5cm} |p{2cm}|p{2cm}|p{1.5cm}|p{1cm}|c|c|c|}    
\hline 
Reference  &  Networks & Scenario   & Use Case & Allocation Method & Allocation Objective & Allocated Parameters  & BS/RSU Assisted & Mobility \\[0.5ex]  
\hline

 \cite{Zheng2013} & LTE, DSRC & Intersection  & Relaying & Hungarian method & Maximizing transmission capacity & Bandwidth   & \cmark &  \cmark  \\
\hline
 \cite{Guo2019_h}  & LTE, DSRC &Two-way urban roadway  & Generic & Hungarian algorithm & Maximizing sum rate & Bandwidth, power   & \cmark &  \cmark  \\
\hline
 \cite{Xiao2018}   & LTE,  TV White Space &Urban roads and intersections  & Generic & Game Theory & Maximizing achievable data rate  & Bandwidth, Power   & \cmark &  \cmark  \\
\hline
 \cite{Huang2018} & LTE, WiFi & Urban layout  & Non-safety applications & Greedy algorithm & Maximizing achievable rate  & Bandwidth  & \cmark &  \xmark  \\
\hline
 \cite{Cao2016}   & Cellular, DSRC & multi-lane highway & Generic & Hungarian method & Minimizing delay & Bandwidth  & \cmark &  \cmark  \\ 
\hline
\end{tabular}

\end{table*}

\subsection{Network Slicing based RA} \label{sec:RA_NS}
\begin{figure*}[!ht]
  \centering
  \includegraphics[trim=0.5cm 0.5cm 0.5cm 0.5cm, clip=true, width=6in]{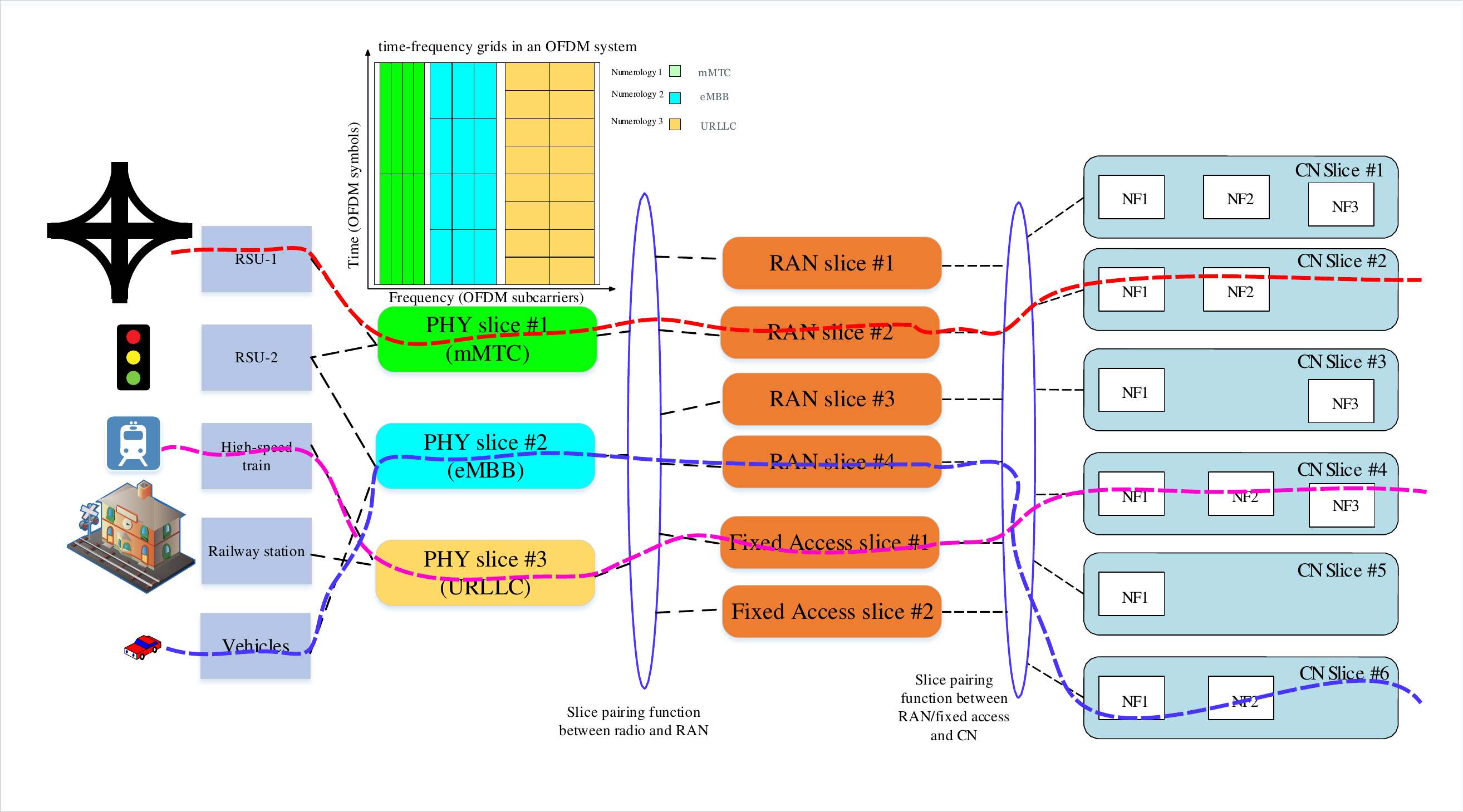}\\
  \caption{Network slicing implemented across different layers (e.g., PHY, RAN, CN) for a C-V2X network consisting of RSUs, high-speed trains, railway stations and moving vehicles.}
  \label{NS}
\end{figure*}

 Network slicing (NS) is a new paradigm that has arisen in recent years which helps to create multiple logical networks on top of a common physical network substrate tailored to different types of data services and business operators \cite{NGMN2015,Khan2020}. NS offers an effective way to meet the requirements of varied use cases and enables individual design, deployment, customization, and optimization of different network slices on a common infrastructure \cite{Foukas2017}.
In addition to providing vertical slices (for vertical industries), NS may be used to generate horizontal slices which aim to improve the performance of User Equipment (UE) and enhance the user experience \cite{Intel2016}. Although initially proposed for the partition of Core Networks (CN), using techniques such as Network Function Virtualization (NFV) and Software Defined Networking (SDN) \cite{Bektas2018}, the concept of NS has been extended to provide efficient end-to-end data services by slicing radio resources in Radio Access Networks (RANs) as well \cite{Sexton2019,Escudero2019}. The slicing of radio resources mainly involves dynamic allocation of time and frequency resources based on the characteristics of multiple data services. This is achieved by providing multiple numerologies, each of which constitutes a set of data frame parameters such as multi-carrier waveforms, sub-carrier spacings, sampling rates, and frame and symbol durations. For example, an mMTC slice in C-V2X is allocated with relatively small subcarrier spacing (i.e., for massive connectivity) and hence large symbol duration. In contrast, URLLC requires large subcarrier spacing to meet the requirements of ultra-low latency and stringent reliability. Fig. \ref{NS} depicts how NS is implemented across different layers (e.g., PHY, RAN, CN) of a C-V2X network consisting of RSUs, high-speed trains, railway stations and vehicles. Using orthogonal frequency-division multiplexing (OFDM) as the transmission scheme, the three types of time-frequency grids (shown in different colors) in Fig. \ref{NS} correspond to the three classes of numerologies for mMTC, eMBB, and URLLC, respectively. Roughly speaking, eMBB and URLLC slices may help address the second major challenge presented in Section I, whereas mMTC slices aim to address the third major challenge. These slices are configured according to specific QoS requirements of various C-V2X use cases.

A step-wise approach for designing and applying function decomposition for NS in a 5G CN has been proposed in \cite{Sama2016}. Their main idea is to identify those functions which could be merged in different network elements as well as their corresponding implications for communication procedure and information storage.
\cite{Soenen2017} presented a concrete example of using NS in the vehicular network domain focusing on efficient notification of 
unexpected road conditions among cars within a certain range. By properly configuring the SDN switch and controller, it is shown in \cite{Soenen2017} that a network slice for such inter-car communication can be readily created. For ultra-low latency in autonomous driving, a scalable and distributed CN architecture with the aid of 5G NC has been proposed to allow the deployments of fog, edge and cloud computing technologies \cite{Chekired2019}. The benefits of 5G NC (in comparison with 4G NC) for efficient C-V2X have been discussed in \cite{Seremet2019}.

In \cite{Silva2016}, the impact of NS on a 5G RAN, such as the CN/RAN interface, the QoS framework, and the management framework, has been discussed. It is pointed out in \cite{Silva2016} that dynamic NS is preferred in order to cater for rapid changes in traffic patterns. Comprehensive work on applications of NS to support a diverse range of C-V2X use cases is presented in \cite{Campolo2017}. Major C-V2X slices identified in \cite{Campolo2017} are: autonomous driving, tele-operated driving, vehicular infotainment, and vehicular remote diagnostics and management. For example, the slice for supporting tele-operated driving enables URLLC and the slice for vehicular infotainment may use multiple Random Access Technologies (RATs) to support higher throughput. \cite{Campolo2017} also show that slicing may be carried out in different vehicular devices according to their storage and computing capacities as well as the nature of the data services, a scenario similar to mobile edge computing.

It is noted that NS can be carried out not only at higher levels of wireless networks, but also in the PHY. In 2017,  a multi-service system framework implemented in both time and frequency domains was proposed \cite{Zhang2017,Zhang-VTC2017}. A major issue here is how to select and design multicarrier waveforms with good time-frequency localization, low out-of-band power emission, low Inter-Carrier Interference (ICI) among different sub-bands using different numerologies, and capability to support multi-rate implementation. Multicarrier waveform design for PHY NS such as Filtered Orthogonal Frequency-Multiple Access (F-OFDMA), windowed-OFDM, and Universal Filtered Multi-Carrier (UFMC) have been studied in \cite{Zhang2017,Zhang2018,Zhang-TVT2018}.

\section{RA for Heterogeneous Vehicular Networks}

A  graph based resource scheduling approach was proposed in \cite{Zheng2013} for cooperative relaying
in heterogeneous vehicular networks. In LTE, vehicles close to the base station usually enjoy high data rates due to favourable radio links, while vehicles far away from the base station suffer from lower data rates due to poor channel conditions. To tackle this problem, cooperative relaying may be adopted to establish V2V communications for distant vehicles through DSRC. \cite{Zheng2013} proposed  a  bipartite graph based scheduling scheme  to determine the transmission strategy for each vehicle user from base station (i.e., cooperative or non-cooperative) and  the selection of relaying vehicles. The scheme proposed in \cite{Zheng2013} consists of the following three steps: 1) construct a  weighted bipartite graph, where the weight of each edge is determined based on the capacity of the corresponding V2V link, 2) solve the maximum weighted matching problem using the Kuhn-Munkres algorithm (also known as Hungarian method) \cite{Munkres1957,Kuhn1955}, and 3) optimize the number of messages that need to be relayed, where  binary search was utilized to find the optimal solution.  The proposed approach guarantees  fairness among vehicle users  and can improve the data rates for the vehicles far away from the base station.

\begin{figure}[htbp]
  \centering
  \includegraphics[width=\figwidth]{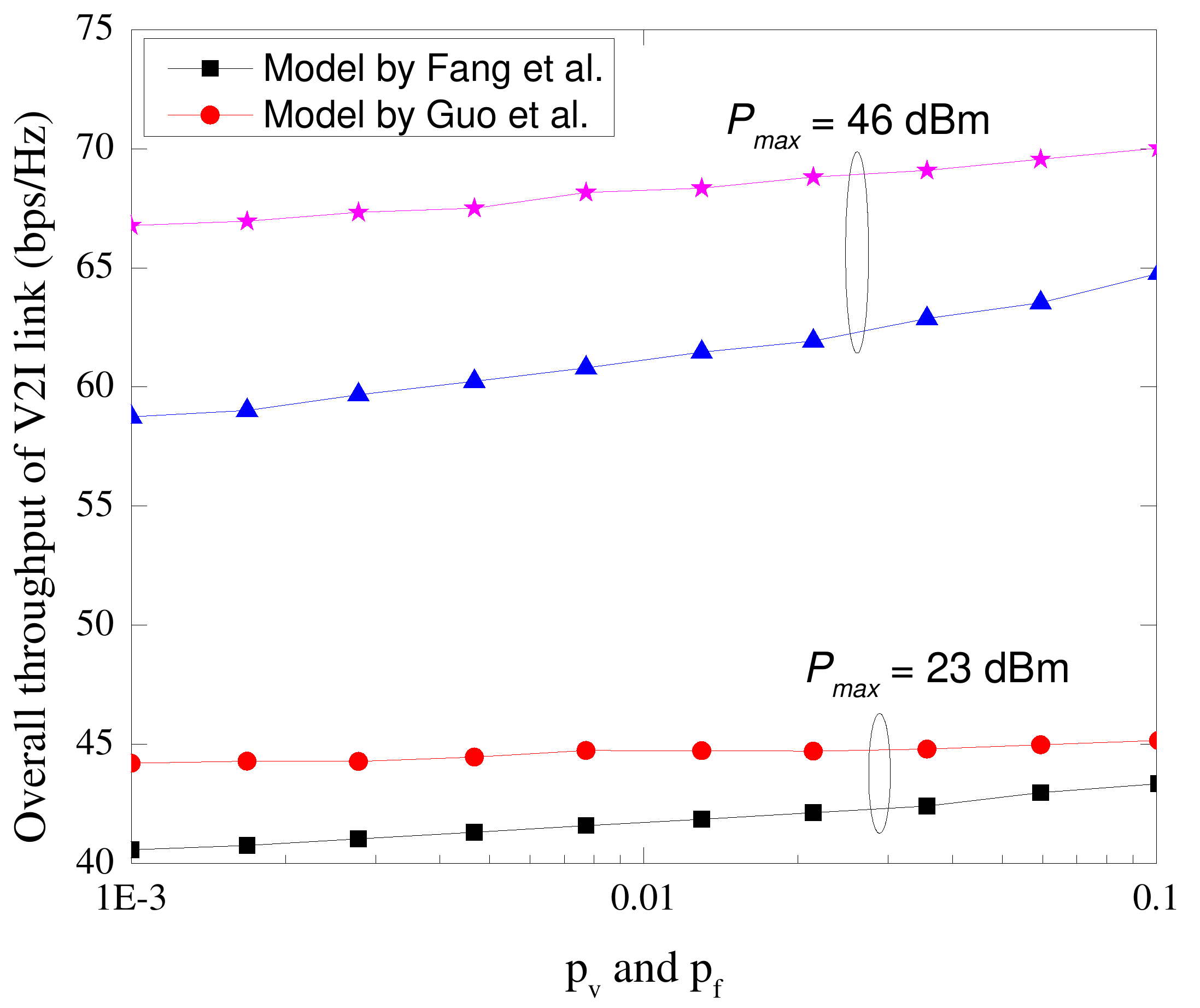}\\
  \caption{Throughput comparison between schemes by Guo et al. \cite{Guo2019_h} and Fang et al. \cite{Fang2017} with respect to reliability of the V2V link $(p_v)$ and cellular user link $(p_f)$ \cite{Guo2019_h}.}\label{fig:Guo2019a}
\end{figure}

\begin{figure}[htbp]
  \centering
  \includegraphics[width=\figwidth]{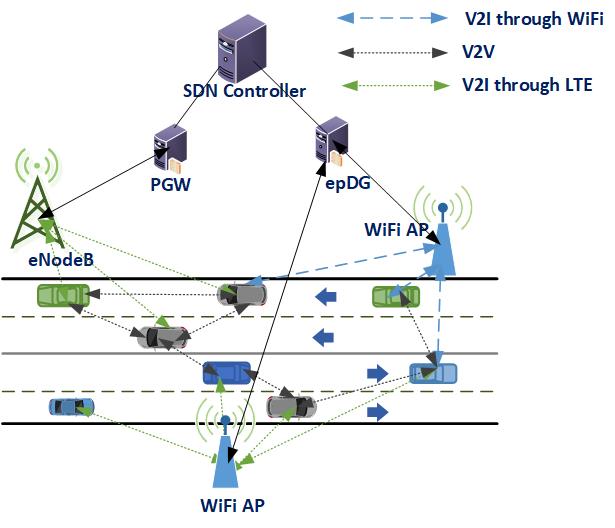}\\
  \caption{Software defined network (SDN) based heterogeneous vehicular network.}\label{fig:Huang2018a}
\end{figure}


Very recently, a cascaded  Hungarian channel allocation algorithm was presented by Guo et al. \cite{Guo2019_h} for non-orthogonal multiple access (NOMA) based heterogeneous vehicular networks.  \cite{Guo2019_h}  addressed the channel assignment problem  in  high-mobility  environments with different user QoS requirements and imperfect CSI by formulating a chance constrained throughput optimization problem.  To validate the proposed model, the authors in  \cite{Guo2019_h}  simulated a two-way urban roadway scenario. The vehicles were covered by a single macro-cell  and  several  non-overlapping  coexisting  femto-cells.  The vehicles positions were based on a spatial  Poisson  point  process and  constant vehicle speed (60 km/h) was considered. In Fig. \ref{fig:Guo2019a},  the overall throughput is compared with that of the RA method reported in \cite{Fang2017}. Enhanced performance is observed for the allocation scheme of \cite{Guo2019_h}, thanks to an efficient user scheduling algorithm which fully utilizes the transmit power to maximize the throughput. It is also observed that the method proposed in \cite{Guo2019_h} provides more benefits with increasing transmit powers.

Xiao et al. \cite{Xiao2018} investigated the spectrum sharing for vehicle users in heterogeneous vehicular networks by exploiting available white space spectrum such as TV white space spectrum.  A non-cooperative game theoretic approach was proposed with correlated equilibrium. Their proposed approach allows macro-cell base stations to share the available spectrum with the vehicle users and improves the spectrum utilization by  reusing the white space spectrum without degrading the macro-cell performance. By sharing available spectrum with LTE and Wi-Fi networks,  \cite{Huang2018} presented a Quality of Experience (QoE) based RA scheme for a software defined heterogeneous vehicular network. The system model considered in \cite{Huang2018} is shown in Fig.~\ref{fig:Huang2018a}.  To maximize the QoE of all vehicular users, the proposed scheme exploits the CSI of vehicular users  to extract  transmission qualities of those users with different access points. A heuristic solution was proposed to allocate the available resources (in LTE and Wi-Fi networks), which can be used in  both centralized and hybrid software defined network systems.  With 20 vehicles, a remote server with an SDN controller, one eNodeB and three Wi-Fi access points, authors in \cite{Huang2018} presented the performance comparison between the proposed SDN based scenario and non-SDN based scenario. In the non-SDN based scenario, the optimization for  the allocation of LTE and Wi-Fi resource is carried out separately. Due to the joint optimization of RA, the proposed  method allocates resources effectively and hence outperforms its non-SDN counterpart. An allocation approach for joint LTE and DSRC networks was proposed in \cite{Cao2016}. The proposed approach allocates the LTE resources to minimize the number of vehicles that compete for channel access in DSRC based communication. The LTE resources are  optimally allocated by the eNodeB, which jointly pairs one vehicle with another and allocates the resources to the pair considering a guaranteed signal strength for all communication links.


\begin{figure*}
	\centering
	\includegraphics[scale=1.0]{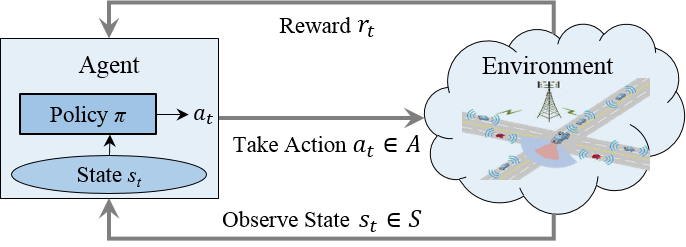}
	\caption{The structure of reinforcement learning for V2V links \cite{Ye2018ICC}.	 In the learning framework, a V2V link (Agent)  learns the policy  to select the  sub-band \& Tx power (Action) considering channel info, the remaining traffic amount, the latency constraint, the interference level (State) and the achieved capacity of V2I links and V2V latency (Reward).}
	\label{fig:sysModelML}
\end{figure*}

\begin{table*}[h]

\centering          

\caption{Existing RA techniques with  machine learning.}  \label{tab:comp_ML}
\scriptsize
\centering          
\begin{tabular}{| p{1cm} | p{1.5cm}| p{3.1cm}| p{2.8cm}| p{3.2cm} |p{2.2cm}|p{0.8cm}|} 
\hline 
Reference  &Learning Technique &  Networks & Scenario   & Allocation Objective & Allocated Parameters  & Mobility
\\
\hline
\cite{Ye2018ICC}  & Multi-agent (deep) RL &  D2D-based V2V communication for safety message
                            & Unicast for multiple vehicles  in intersection   & Min. of interference and guarantee the latency const.
                            & Sub-band \& Tx power   & \cmark \\
\hline
\cite{Ye2019}  & Multi-agent (deep) RL &  D2D-based V2V communication for safety message
                       & Unicast \& Broadcast  in intersection  & Min. of interference and guarantee the latency const.
                        & Sub-band and messages to broadcast   & \cmark \\
\hline
\cite{Pressas2017}  & RL &  DSRC-based V2V communication & Broadcasting from multiple vehicles
                              & Reduction of the packet collision and bandwidth waste  & Contention window size  & \xmark  \\
\hline
\cite{Li2017-2}  & RL &  V2I in a Hetnet (macro, femto, \&  pico) & Downlink transmissions     & Load balancing  & User association & \cmark \\
\hline
\cite{Xu2014}  & RL   &  V2I in a Hetnet (cellular \& DSRC)  & (Non-real time) infotainment data downlink
                       & Seamless mobility management  & Handoff decisions  & \cmark   \\
\hline
\cite{Salah2016}  & MDP based RL  &  Vehicular cloud supporting multiple service types  & Dynamic change of the required QoS level
                            & Efficient resource utilisation  & Network resource  & \xmark   \\
\hline
\cite{Zheng2016}  & POMDP based RL &  Hetnets with multiple (cellular-based) virtual BSs
                             & 400m long highway with two virtual BSs supporting  multiple vehicles   & Efficient resource utilisation
                             & Virtualised radio resource block  & \cmark
 \\[0.5ex]  
\hline

\hline
\end{tabular}

\end{table*}

\section{Machine Learning based RA for Vehicular Communications }
 \label{Sect:MLRA}

In vehicular networks, whilst vehicles are expected to employ various facilities such as advanced on-board sensors including radar and cameras and even high-performance computing and storage facilities, massive amounts of data will be generated, processed and transmitted. Machine Learning (ML) is envisaged to be an effective tool to analyse such a huge amount of data and to make more data-driven decisions to enhance vehicular network performance \cite{Liang2019}. For details on machine learning, readers can refer to \cite{Jiang2017,Sutton1998,Arulkumaran2017}.


For resource allocation, the traditional approach is to formulate an optimisation problem and then obtain an optimal or sub-optimal solution depending on the trade-off between target performance and complexity. However, in vehicular networks where the channel quality and network topology can vary continuously, the conventional optimization approach would potentially need to be rerun whenever a small change happens, thus incurring huge overhead \cite{Ye2018}. While an ML approach could be an alternative to prevalent optimisation methods, research on applying ML in vehicular networks is still at an early stage \cite{Liang2019}.
In the existing literature \cite{Ye2018ICC, Ye2019, Pressas2017, Li2017-2,Xu2014,Salah2016,Zheng2016},
machine learning has been applied to resource (e.g., channel and power) allocation, user association, handoff management, and virtual resource management for V2V and V2I communications while considering the dynamic characteristics of a vehicular network.

A distributed channel and power allocation algorithm employing deep reinforcement learning (RL) \cite{Arulkumaran2017}  has been proposed for cellular V2V communications  in \cite{Ye2018ICC}. With the assumption that an orthogonal resource is allocated for V2I links beforehand, the study focuses on resource allocation for V2V links under the constraints of V2V link latency and minimized  interference impact to V2I links. The structure of reinforcement learning for V2V links is shown in Fig. \ref{fig:sysModelML}. While the agent corresponds to each V2V link, it interacts with the environment which includes various components outside the V2V links. The state for characterising the environment is defined as a set of the instantaneous channel information of the V2V link and V2I link, the remaining amounts of traffic, the remaining time to meet the latency constraints, and the interference level and selected channels of neighbours in the previous time slot. At time epoch $t$, each V2V link, as an agent, observes a state $s_t \in \mathcal{S}$, and depending on its policy $\pi$, takes an action $a_t  \in \mathcal{A}$, where $\mathcal{S}$ is the set of all states and $\mathcal{A}$ the set of all available actions. An action refers to the selection of the sub-band and transmission power. Following the action, the agent receives a reward $r_t$ calculated by the capacity of V2I links and the V2V latency. The optimal decision policy $\pi$ is determined by deep learning.

The training data is generated from an environment simulator and stored. At the beginning, for the training stage, the generated data is utilised to gradually improve the policy used in each V2V link for selecting spectrum and power. Then, in a test stage, the actions in V2V links are chosen based on the policy improved by trained data.
\color{black}
This work is extended in \cite{Ye2019} to include a broadcast scenario.
In \cite{Ye2019}, each vehicle is modelled as an agent and the number of times that the message has been received by the vehicle and the distance to the vehicles that have broadcast are additionally considered in defining the state. Then, each vehicle improves the  messages broadcast and  sub-channel selection policies through the  learning mechanism.

%

\begin{figure*}
	\centering
	\begin{subfigure}[b]{0.5\textwidth}
	 \centering
		\includegraphics[trim=5cm 10cm 0.5cm 10cm, clip=true, width=4.7in]{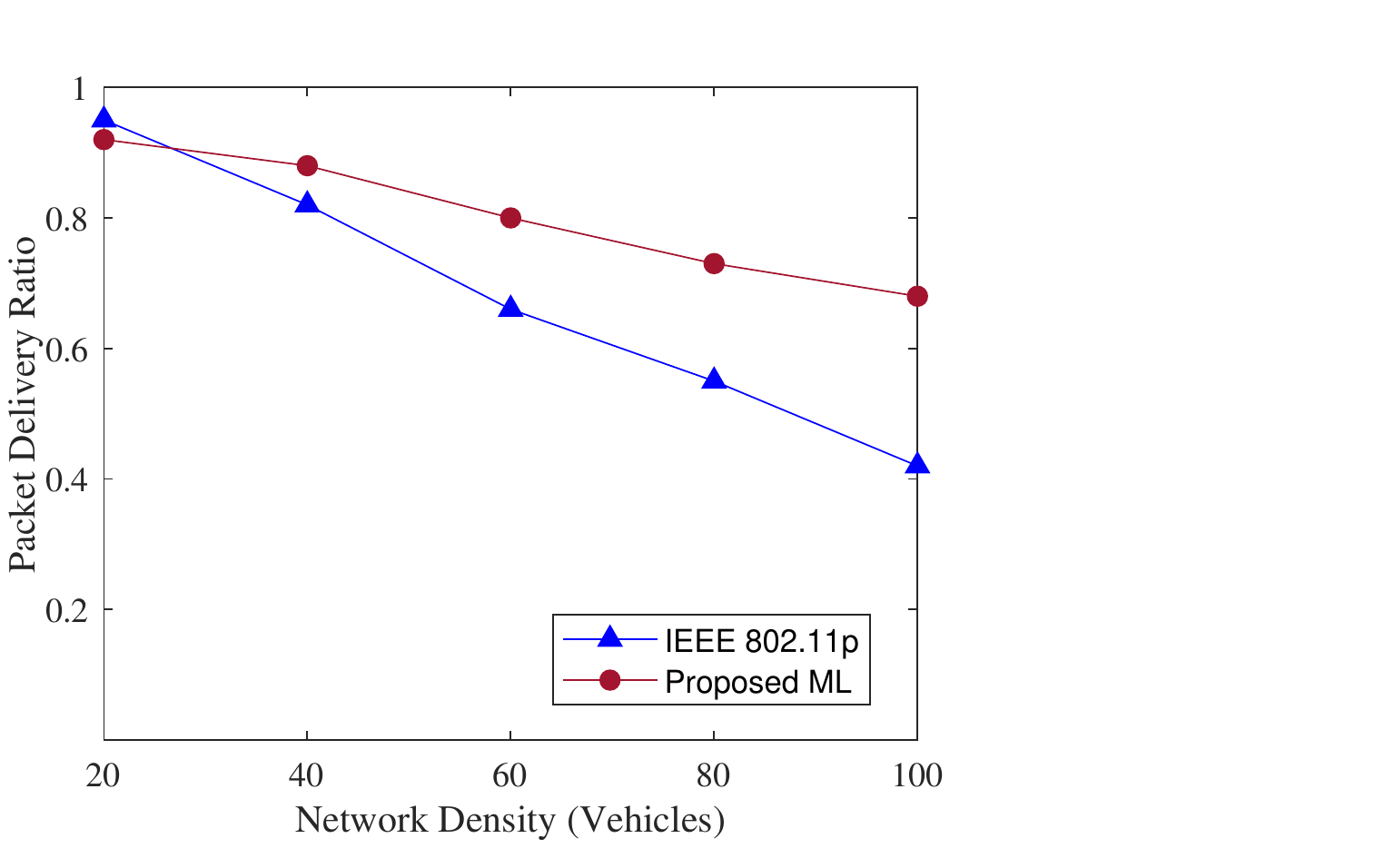}
		\caption{Packet delivery ratio vs. network density for 256 byte packets}
		\label{fig:Pressas2017a}
	\end{subfigure}~
	\begin{subfigure}[b]{0.5\textwidth}
	 \centering
		\includegraphics[trim=5cm 10cm 0.5cm 10cm, clip=true, width=4.7in]{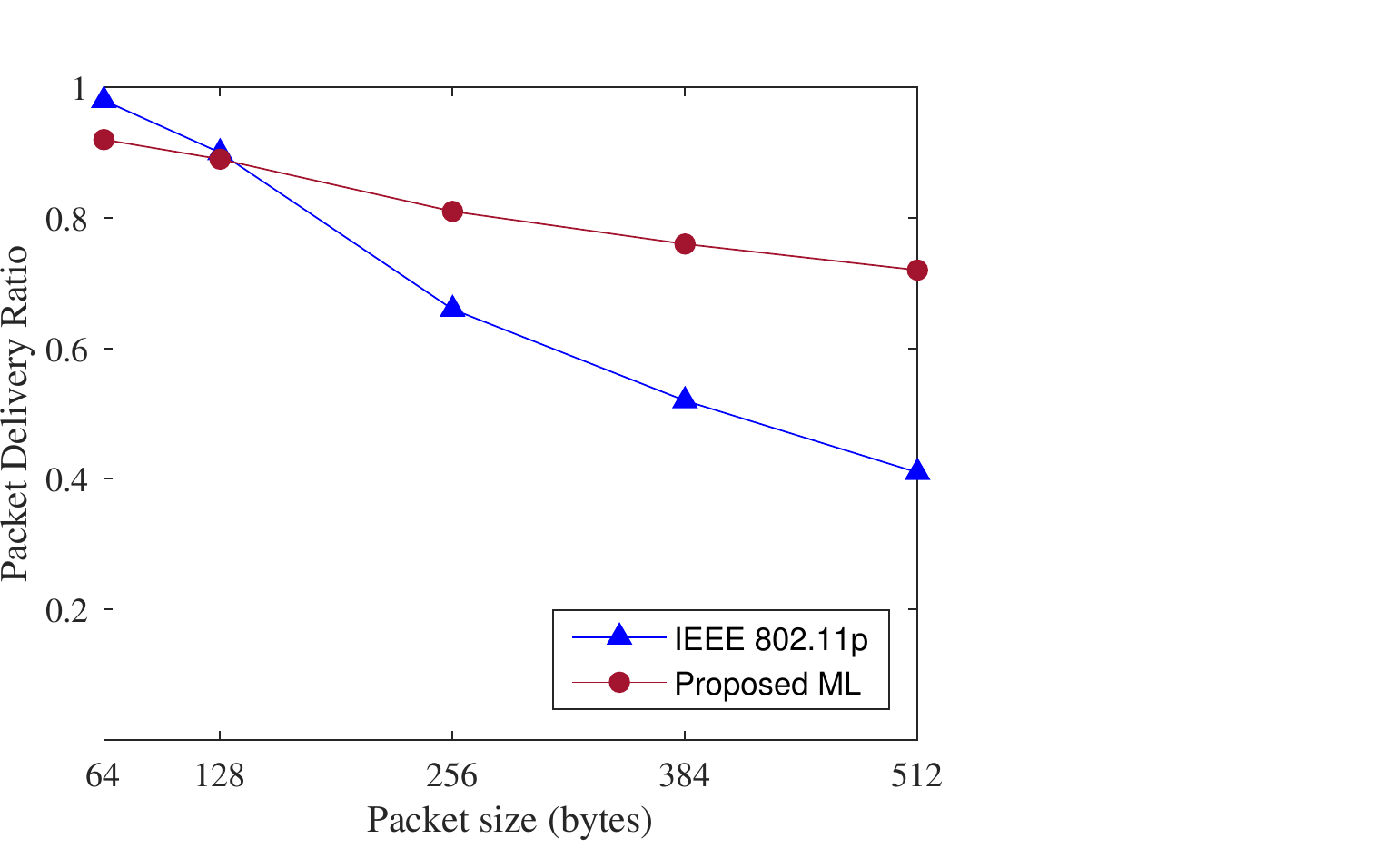}
		\caption{Packet delivery ratio vs. packet size  for 60 vehicles}
		\label{fig:Pressas2017b}
	\end{subfigure}%
	\caption{Performance comparison between IEEE 802.11p standard and the proposed ML based approach for DSRC \cite{Pressas2017}.}\label{fig:Pressas2017}
\end{figure*}

In \cite{Pressas2017}, a contention-based MAC protocol for V2V broadcast transmission
using the IEEE 802.11p standard for DSRC is investigated.
In a scenario with fewer than 50 vehicles,
IEEE 802.11p can exhibit better performance than LTE in terms of lower latency and higher packet delivery ratio than LTE. However, as vehicle density gets high, the standard becomes unable to accommodate
the increased traffic.
In \cite{Pressas2017}, with the aim of overcoming the scalability issue associated with the vehicular density,
an ML based approach is proposed to find the optimal contention window to enable efficient data packet exchanges with strict reliability requirements.
As a independent learning agent, each vehicle employs learning to decide on the contention window size.
The result of each packet transmission, either success or fail, is feedback
and utilized for the window size decision. Similar to \cite{Ye2018ICC}, the two-stage RL is considered to get instant performance benefits starting from the first transmission. At the beginning, the  data generated from a simulator is exploited to  improve the policy. In the test stage, the actions are chosen based on the pre-trained policy while the policy keeps improving.  Authors in \cite{Pressas2017} evaluated  the performance of their proposed ML based approach via  simulation. Through simulation results illustrated in Fig.~\ref{fig:Pressas2017},  it is shown that the proposed ML based approach achieves more reliable packet delivery and higher system throughput performance.  In the simulation, all cars in the area of $600 m \times 500 m$ are assumed to continuously transmit broadcast packets with a period 100 ms. While the packets are transmitted using the highest priority, the network density  changes. In  Fig.~\ref{fig:Pressas2017a} for a given packet size 256 bytes, it is shown that the proposed approach reduces collisions between data packets and achieves better packet delivery ratio (PDR) performance in denser networks by adjusting the size of contention window. In a sparse networks (of 20 cars), while a minimum window size is optimal, the learning protocol exploring larger window size causes increases of packet collisions. However, In denser networks, the proposed approach is superior to IEEE 802.11p standard. In a network formed of 80 cars, a 37.5\% increase in PDR performance is observed. In  Fig.~\ref{fig:Pressas2017b}, the performance of the proposed algorithm is evaluated for different packet sizes in a network of 60 cars. While the proposed approach achieves more reliable packet delivery, it yields 72.63\% increases in throughput for 512 bytes packet size.

In \cite{Li2017-2},
the ML approach is exploited to develop the user association algorithm for load balancing in heterogeneous vehicular networks. Considering   data flow (generated from vehicular networks) characteristics in the spatial-temporal dimension, a two-step association algorithm is proposed. The initial association decision is made by a single-step
reinforcement learning approach
\cite{Sutton1998}. Subsequently,
 a \color{black}
base station (i.e., macro, pico and femto cells) uses historical association patterns to make decisions for association. In addition, a base station, as an agent of learning, keeps accumulating feedback information and updates the association results adaptively. While each base station runs the proposed algorithm in a distributed manner, in the long run, it is shown that both the real-time feedback and the regular traffic association patterns help the algorithm deal with the network changes.

In \cite{Xu2014}, a vertical handoff strategy has been devised by using a fuzzy Q-learning approach \cite{Sutton1998} for heterogeneous vehicular networks consisting of a cellular network with global coverage complemented by the V2I mode. From the OBU side, various information including  average Received Signal Strength (RSS) level, vehicle velocity and  the type of data
is sent to the RSU side. Then, the RSU side  considers the delivered information as well as the traffic load (i.e., the number of users associated with the target network) and makes handoff decisions by using the fuzzy Q-learning method. With the simulation results, it is shown that the proposed algorithm, which has a real-time learning capability, can determine the network connectivity to ensure seamless mobility management without prior knowledge of handoff behaviour.

In \cite{Salah2016,Zheng2016}, a machine learning approach is exploited to devise the virtual resource allocation in vehicular networks. Vertical clouds \cite{Lee2014} consisting of various OBUs, RSUs, and remote cloud servers can provide a pool of processing, sensing, storage, and communication resources that can be dynamically provisioned for vehicular services.
The importance of resource allocation in the vehicular cloud is highlighted in \cite{Salah2016}. Poorly designed resource allocation mechanisms could result in QoS violation or under-utilisation of resources, whereas dynamic resource provisioning techniques are crucial for meeting the dynamically changing QoS demands of vehicular services. Against this background, a reinforcement learning framework has been proposed for resource provisioning to cater for dynamic demands of resources with  stringent QoS requirements.
In  \cite{Zheng2016}, a two-stage delay-optimal dynamic virtualisation radio scheduling scheme has been developed. Based on the time-scale, the proposed algorithm is divided into two stages, macro allocation for large time-scale variables (traffic density) and micro allocation with short time-scale variables (channel state and queue state). The dynamic delay-optimal problem is formulated as a partially Observed Markov Decision Process (POMDP) \cite{Jiang2017} and is then solved by an online distributed learning approach.


In Table \ref{tab:comp_ML}, the characteristics of ML based algorithms in literature are summarised.
Since the increase of communication overheads and the computational complexity to analysis a high volume of data can significantly deteriorate the performance of vehicular networks,
aforementioned works  consider a distributed learning approach.
Different entities are chosen as a autonomous agent to manage their problem: a V2V link in \cite{Ye2018ICC,Ye2019}, a vehicle in \cite{Pressas2017}, a BS in \cite{Li2017-2}, a RSU in \cite{Xu2014}, and resource controller \cite{Zheng2016}.
In \cite{Salah2016}, whilst it focuses on the benefit of the learning-based dynamic resource provisioning, a learning framework is considered.

In machine learning, the type of data (i.e., labelled or unlabelled) can be a key element to decide the  learning technique to use and
high-quality data is an important factor in affecting the learning performance.
However, the scarcity of real datasets available for vehicular networks is pointed out as one of the biggest challenges for the application of machine learning \cite{M-Cayamcela2019}.
Different from learning approaches requiring datasets obtained in advance (i.e., supervised, unsupervised learning),
the RL approach can be exploited without prior knowledge of the environment.
In the aforementioned studies, the RL approach is exploited without any prior datasets and it is shown that online RL approach can converge to a solution through feeding back from the dynamic vehicular environment iteratively.
\color{black}

\section{Future Research Directions} \label{Sect:Future_Dir}

In this section, we present a number of attractive directions for future research in resource allocation for vehicular networks.
\subsection{RA for NR-V2X and IEEE 802.11bd}
While NR-V2X is emerging as an improved version of LTE-V2X, the IEEE 802.11bd standard has recently emerged as an upgraded version of the IEEE 802.11p standard to reduce the gap between DSRC and C-V2X \cite{Naik2019}. Both of the upgraded technologies are expected to support mm-Wave communications, which raise one of the main challenges, that of effective utilization of traditional bands and new mm-Wave bands. As such, suitable dynamic resource scheduling is required to exploit their unique benefits. For example, while  mm-Wave communication offers  very high data rates, it is mostly suitable for short-range  communication. Thus, the resource allocation approach should allocate resources in mm-Wave bands to those transmitters with receivers within short range. For the out-of-coverage scenario, NR-V2X has introduced  co-operative distributed scheduling approaches, where vehicles can either assist each other in determining the most suitable transmission resources or a vehicle  schedules the sidelink transmissions for its neighboring vehicles. In the first scenario,  a thorough investigation is required to determine the type of information (e.g., packet reception acknowledgment, channel busy ratio assessment, etc.) that vehicles need to share to improve the resource allocation process, while ensuring that the sharing process itself will not cause congestion in the vehicular networks. On the other hand, the autonomous selection of a cluster-head (a vehicle that allocates the resources for its surrounding vehicles) is an open issue for the latter scenario. For example, what information shared by vehicles benefits the nomination of a cluster-head, how to adapt cluster-head selection algorithms to different vehicular environments (e.g., highway, intersection, urban, rural, etc.), while ensuring good connectivity between the cluster-head and other vehicles.


\subsection{Efficient and Ultra-Fast Slicing for C-V2X}
For NS discussed in Subsection~\ref{sec:RA_NS}, it is critical to understand how C-V2X competes for system resources with other vertical applications, how C-V2X assigns and optimizes these  resources among a vast range vehicular use cases, and in particular, how to carry out efficient and ultra-fast NC in highly dynamic and complex vehicular environments. In a high mobility channel, for example, the PHY slicing for multiple numerologies needs to rapidly deal with severe ICI and inter-symbol interference. An interesting future direction is to design intelligent slicing algorithms by efficiently using various computation resources at the edge or in the cloud. Recent advances on this topic can be found in \cite{Albonda2019,Mei2019,Xiong2019}.

\subsection{Security Enhancement with Blockchain Technology}
The widespread deployment of V2X networks very much relies on significantly enhanced security for large scale vehicular message dissemination and authentication. The consideration for this imposes new constraints for RA in V2X networks. For example, mission critical messages should have ultra resilient security to deal with potential malicious attacks or jamming, whilst multimedia data services prefer lightweight security due to large amount of data rates. These two types of security lead to different frame structures, routing/relaying strategies, and power/spectrum allocation approaches.  
Besides the approaches introduced in Subsection~\ref{sec:RA_Security_CV2X}
it is interesting to investigate the applications of blockchain which has emerged recently as a disruptive technology for secured de-centralized transactions involving multiple parties. An excellent blockchain solution (e.g.,  smart contract or consensus mechanism) should not only allow access to the authenticity of a message, but also preserve the privacy of the sender \cite{Kang2019,Yazdinejad2019}.

\subsection{Machine Learning supported Resource Allocation}

While the potential of applying ML in vehicular networks has been discussed in Section \ref{Sect:MLRA}, mechanisms as to how to adapt and exploit ML to account for the particular characteristics of vehicular networks and services remains a promising research direction.
Vehicular networks significantly differ from the scenarios where machine learning has been conventionally exploited in terms of strong dynamics in wireless networks, network topologies, traffic flow, etc. How to efficiently learn and predict such dynamics based on historical data for the benefit or reliable communications is still an open issue.  
In addition, data is supposed to be generated and stored across various units in  vehicular networks, e.g., OBUs, RSUs, and remote clouds. It could be interesting to investigate whether traditional centralised ML approaches can be exploited to work efficiently in a distributed manner. For collective intelligent decision making in learning-capable vehicular networks, the overhead for information sharing and complexity of learning algorithms
need to be taken into account.


\subsection{Context Aware Resource Allocation}
Existing work on resource allocation for vehicular networks mostly deals with efficient allocation of resource blocks such as frequency carriers or time-slots. However, most of the prior work on resource allocation did not consider context-aware/on-demand data transfer  applications  in vehicular networks. Since on-demand data transfer applications need to meet constraints such as deadline of the requested data items or priority of data items, to ensure a reliable service, there is a need for research to consider those more thoroughly. Although there is a lot of prior work \cite{Zhan2011,Wang2014}  on performance evaluation of on-demand data dissemination scenarios in terms of the above constraints, they do not deal with the allocation of resource blocks, which is important for 5G networks.

\section{Conclusions} \label{Sect:Conclusions}
In this paper, we have surveyed  radio resource allocation schemes in vehicular networks.  We have categorized these schemes into three categories  based on the types of  vehicular networks, i.e., DSRC vehicular networks, cellular vehicular networks, and heterogeneous vehicular networks. For each category, the available literature on resource allocation is  reviewed and summarized while highlighting the advantages and disadvantages of the reviewed schemes. We have also discussed several open and challenging future research directions for radio resource allocation in vehicular networks. It is anticipated that this paper will provide a quick and comprehensive understanding of the current state of the art in radio resource allocation strategies for vehicular networks while attracting and motivating more researchers into this interesting area.

\section{acknowledgement}
The authors are deeply indebted to the Editor and the anonymous Reviewers for many of their insightful comments which have greatly helped improve the quality of this work.

\bibliography{refs}
\bibliographystyle{IEEEtran}

\end{document}